\documentclass[10pt]{article}
\usepackage{amsmath}
\usepackage{graphicx}
\usepackage{color}
\usepackage{orcidlink}
\usepackage{hyperref}
\usepackage{multirow}
\usepackage{rotating}
\usepackage{multicol}
\hypersetup{colorlinks=true, linkcolor=blue, citecolor=blue, urlcolor=blue}
\usepackage{caption}
\usepackage{subcaption}
\usepackage{setspace}
\RequirePackage[numbers,sort&compress]{natbib}
\begin{document}
\baselineskip=18pt

\begin{center}
\LARGE{Static AdS Black Holes Surrounded by Strings and Quintessence-like Field within  Rastall Gravity Framework }
\par\end{center}

\vspace{0.3cm}

\begin{center}
{\bf Allan. R. P. Moreira\orcidlink{0000-0002-6535-493X}}\footnote{\bf allan.moreira@fisica.ufc.br }\\ 
{\tt Secretaria da Educa\c{c}\~{a}o do Cear\'{a} (SEDUC), Coordenadoria Regional de Desenvolvimento da Educa\c{c}\~{a}o (CREDE 9),  Horizonte, Cear\'{a}, 62880-384, Brazil}\\
\vspace{0.1cm}
{\bf Faizuddin Ahmed\orcidlink{0000-0003-2196-9622}}\footnote{\bf faizuddinahmed15@gmail.com}\\
{\tt Department of Physics, Royal Global University, Guwahati, 781035, Assam, India}\\
\vspace{0.1cm}
{\bf Abdelmalek Bouzenada\orcidlink{0000-0002-3363-980X}}\footnote{\bf abdelmalekbouzenada@gmail.com (Corresp. author)}\\ 
{\tt Laboratory of Theoretical and Applied Physics, Echahid Cheikh Larbi Tebessi University 12001, Algeria}\\
\vspace{0.1cm}

\end{center}

\vspace{0.3cm}

\begin{abstract}
In this work, we tested the physical properties of Rastall black holes in the presence of string clouds and quintessence. The modified spacetime geometry arising from the Rastall framework is first established, providing the basis for analyzing the geodesic structure. We study null geodesics in detail, testing photon trajectories, the conditions for photon spheres, BH shadows, the associated effective radial force, and the topological features of photon rings. Also, the analysis is further tested to timelike geodesics, where we investigate the motion of massive particles and determine the innermost stable circular orbits. In this case, we discuss the thermodynamic behavior of the system, show the effects of Rastall gravity, strings, and quintessence on the BHs stability and thermal characteristics. In this context, our results improve and show of how modifications to general relativity and surrounding matter distributions influence both the dynamical and thermodynamical aspects of BHs.
\end{abstract}

\vspace{0.3cm}

{\bf Keywords:} { Rastall gravity; black holes; string clouds; quintessence; geodesics; photon sphere; black hole shadow; thermodynamics.}

\section{Introduction}\label{intro}

The investigation of black holes (BHs) in Anti-de Sitter (AdS) spacetimes has gained remarkable importance due to the celebrated AdS/CFT correspondence, which establishes a profound link between gravitational theories in AdS geometries and conformal field theories defined on the boundary \cite{BHR1,BHR2,BHR3,BHR4}. Another important result is that this duality not only provides a conceptual framework for illustrating quantum gravity (QG) but also serves as a valuable tool for studying strongly coupled field theories and their thermodynamic properties. A further extension of these studies arises from the incorporation of cosmic string (CS) distributions surrounding the BH, leading to nontrivial modifications of both the metric and thermodynamic quantities of the system \cite{BHR5,BHR6,BHR7,BHR8}. The introduction of such string configurations can be traced back to Letelier’s pioneering models, where an ensemble of one-dimensional ($1\mathcal{D}$) defects is used to describe the effective influence of strings on the background geometry \cite{BHR9,BHR10,BHR11,BHR12,BHR13}. In cosmological contexts, these strings are understood as possible relics of early-universe symmetry-breaking phase transitions, and they may contribute significantly to the explanation of cosmic structure formation \cite{BHR14,BHR15}. The study of a CS parameter in BH models modifies several fundamental properties, such as the location of the horizon, the geodesic structure of test particles, and the thermal stability ($T$) of the system \cite{BHR16,BHR17}. In this context, the presence of CS in AdS BH models provides a versatile platform for illustrating deviations from classical BH physics and testing the influence of gravitational theories in both classical and quantum regimes \cite{BHR18}. 

The investigation of geodesic motion of both massive and massless particles in BH geometries has always been central in understanding the astrophysical implications of strong gravity. Such studies provide more information about relativistic regimes where Newtonian descriptions fail. In this case, the groundbreaking detection of gravitational waves (GW) produced by binary BH mergers \cite{PD1} has further reinforced the importance of exploring the dynamics of particles and photons near compact objects. Through particle trajectories, one can probe several relativistic effects, including perihelion precession, bending of light, gravitational lensing, quasi-periodic oscillations, and gravitational time delay \cite{PD2,PD3}. In particular, timelike circular orbits help to describe the distribution and stability of matter in accretion disks, whereas null circular orbits, associated with the photon sphere, are key to understanding strong lensing phenomena, which not only distort images of background sources but may also amplify their brightness. Also, radial null and timelike geodesics are of great relevance, since they shed light on mechanisms responsible for the generation of relativistic jets from BH environments. Extensive research has been carried out on this subject, with numerous works testing the detailed analysis of timelike and null motion in various spacetimes \cite{PD4,PD5,PD6,PD7,PD8,PD9,PD10,PD11,PD12,PD13}. Also, several studies have considered the motion of neutral and charged particles in the surroundings of both static and rotating BHs, including cases with charge 
 ($Q$) \cite{PD14,PD15,PD16,PD17,PD18,PD19}. In this context, the influence of different external fields, such as magnetic fields, quintessence, perfect fluids, and string clouds, has been systematically examined in relation to the motion of massive and massless test particles around BHs \cite{PD20,PD21,PD22,PD23,PD24,PD25,PD26,PD27}. Beyond the framework of Einstein’s theory, geodesic analyses have also been extended to modified and alternative theories of gravity, to constrain their parameters through observationally relevant predictions \cite{PD28,PD29,PD30,PD31,PD32,PD33,PD34,PD35}. Therefore, the study of particle dynamics and photon orbits not only deepens our understanding of astrophysical processes in strong-field regimes but also serves as a testing ground for the validity of GR and its possible extensions. 

The study of BH thermodynamics stands as one of the most profound achievements in modern theoretical physics, since it intertwines the principles of GR, quantum field theory (QFT), and statistical mechanics (SM) into a unified framework to show and illustrate the nature of gravity at a fundamental level. BHs not only act as gravitational laboratories for testing the interplay between classical and quantum physics but also provide unique insight into the microscopic degrees of freedom that might underlie spacetime itself. Of particular importance are BHs embedded in asymptotically Anti-de Sitter (AdS) geometries, which are central to the celebrated AdS/CFT correspondence. This correspondence, regarded as one of the most powerful dualities discovered in theoretical physics, asserts an exact equivalence between a gravitational theory defined in an AdS bulk and a conformal field theory (CFT) formulated on its boundary \cite{RTH1,RTH2,RTH3}. As a result, thermodynamic investigations of AdS BHs play a dual role: on the gravitational side, they shed light on horizon dynamics and quantum corrections, while on the field-theoretic side, they allow the analysis of strongly coupled many-body systems. In this context, the thermodynamic interpretation of BH parameters becomes essential. The ADM mass of the BH is identified with internal energy, the surface gravity at the event horizon naturally corresponds to temperature, and the area of the horizon encodes entropy \cite{RTH4,RTH6,RTH7,RTH8,RTH9}. These identifications are made precise through the famous relations $S=A/4$, where $A$ denotes the horizon area, and $T=\kappa / 2\pi$, with $\kappa$ being the surface gravity. Such a mapping of geometric to thermodynamic quantities builds a consistent framework in which phase structures, local stability, and global critical behavior can be systematically explored \cite{RTH10}. A landmark discovery in this setting is the Hawking–Page transition: a first-order phase transition between a thermal AdS vacuum and an AdS BH configuration. This process, besides its gravitational interpretation, acquires a holographic meaning in the dual CFT as a transition of confinement/deconfinement phases of gauge fields \cite{RTH11,RTH12}. The onset of such transitions is encoded in non-analytic features of thermodynamic response functions \cite{RTH13,RTH14}, where the heat capacity $ C = T \, \frac{\partial S}{\partial T}, $ often diverges near critical points, signaling the change of stability and the presence of new equilibrium states \cite{RTH15}. Also, to refine the analysis of these transitions and the microscopic interpretation of BH thermodynamics, geometric frameworks have been proposed. Weinhold introduced a Riemannian structure based on the Hessian of the internal energy with respect to extensive variables \cite{RTH16,RTH17}, while Ruppeiner extended this idea by constructing a metric from the Hessian of entropy, closely linked to fluctuation theory in statistical mechanics \cite{RTH18,RTH19}. These two approaches are conformally related, with the conformal factor identified as the inverse temperature \cite{RTH20}. The associated scalar curvature derived from these thermodynamic metrics encodes valuable information about the nature of microscopic interactions: it diverges at phase transitions, distinguishes attractive from repulsive microscopic correlations, and highlights the presence of stability boundaries. Thus, thermodynamic geometry illustrates another mirror of statistical mechanics while simultaneously revealing features that are unique to gravitational systems. In this case, these perspectives not only elevate BHs in AdS backgrounds as crucial testing grounds for ideas in QG but also emphasize their role in bridging gravitational physics, quantum field theory, and statistical mechanics. The combination of BH models, thermodynamic analogies, and geometric methods develops our explanation and illustration of BH microstructure, the dynamics of phase transitions in curved backgrounds, and the possible emergence of a consistent description of quantum spacetime.  

Our study tests the physical characteristics of Rastall BHs immersed in a background of string clouds and quintessence, illustrating how deviations from GR affect their geometry and dynamics. In this case, we begin by formulating the modified spacetime within the Rastall framework, which serves as the foundation for analyzing the motion of test particles.  Also, the null geodesic structure is examined in depth, where we investigate photon trajectories, the criteria for photon sphere formation, the shadow profile of the BH, the effective radial force, and the topology of photon rings. Extending the analysis to timelike geodesics, we focus on the dynamics of massive particles, with particular attention to the ISCOs parameters. In this context, we are testing the thermodynamical properties of the system, discussing how the interplay between Rastall gravity, string clouds, and quintessence influences BH stability, phase behavior, and thermal features. 

The paper is organized as follows: Section (\ref{intro}) introduces the motivation and background of the study. In Section (\ref{sec:2}), the spacetime geometry of Rastall BHs surrounded by strings and quintessence is presented, with emphasis on the modifications arising from the Rastall framework. Section (\ref{sec:3}) is devoted to the geodesic structure, where null geodesics are analyzed in detail; this includes the photon trajectory, the conditions for the photon sphere and BH shadow, the effective radial force, and the topological properties of photon rings. In this case, the discussion then extends to timelike geodesics, focusing on the dynamics of massive particles and the characterization of innermost stable circular orbits (ISCOs). Section (\ref{sec:4}) examines the thermodynamic properties of the BH and illustrates how Rastall gravity, strings, and quintessence affect its thermal behavior and stability. In this context, Section (\ref{sec:5}) shows and explains our results.

\section{Rastall Black Holes Surrounded by Strings and Quintessence }\label{sec:2}

On the other hand, Ref.~\cite{Sarkar:2025fjx} studied charged black holes coupled to a quintessence field within the Rastall gravity framework. It was shown that the quintessence contribution in this scenario enhances the deflection angle compared to the Schwarzschild or Reissner–Nordström (RN) cases. The corresponding charged space-time is described by
\begin{eqnarray}
    ds^2 = -\left(1 - \frac{2M}{r} + \frac{Q^{2}}{r^{2}} 
    - \frac{\mathcal{N}}{r^{\frac{1 + 3\omega - 6\lambda (1+\omega)}{1 - 3\lambda (1+\omega)}}}\right) dt^2+ \left(1 - \frac{2M}{r} + \frac{Q^{2}}{r^{2}} 
    - \frac{\mathcal{N}}{r^{\frac{1 + 3\omega - 6\lambda (1+\omega)}{1 - 3\lambda (1+\omega)}}}\right)^{-1} dr^2 + r^2 \left(d\theta^2+\sin^2\theta\,d\phi^2\right), \label{aa2}
\end{eqnarray}
where $\mathcal{N}$ and $\omega$ are the quintessence field parameters, $\lambda$ encodes the deviation from standard Einstein gravity in the Rastall framework, \(M\) is the black hole mass, and \(\Lambda\) denotes the cosmological constant.

The Letelier BH with a cosmological constant, surrounded by a quintessence-like field (QF), was discussed in Ref.~\cite{MMDC}. The metric reads
\begin{eqnarray}
    ds^2 = -\left(1 - \alpha - \frac{2 M}{r} - \frac{\mathcal{N}}{r^{3w+1}} - \frac{\Lambda}{3} r^2 \right) dt^2 + \left(1 - \alpha - \frac{2 M}{r} - \frac{\mathcal{N}}{r^{3w+1}} - \frac{\Lambda}{3} r^2 \right)^{-1} dr^2 + r^2 \left(d\theta^2+\sin^2\theta\,d\phi^2\right). \label{aa2aa}
\end{eqnarray}

Motivated by these results, we introduce the line element of an asymptotically AdS black hole surrounded by a cloud of strings and a quintessence field in the Rastall framework:
\begin{equation}
     ds^2 = -f(r)\,dt^2 + \frac{dr^2}{f(r)} + r^2 \left(d\theta^2 + \sin^2\theta\,d\phi^2\right), \label{metric}
\end{equation}
where the metric function $f(r)$ is defined as
\begin{equation}
     f(r) = 1 - \alpha - \frac{2 M}{r} - \frac{\mathcal{N}}{r^{\frac{1 + 3\omega - 6\lambda (1+\omega)}{1 - 3\lambda (1+\omega)}}} - \frac{\Lambda}{3} r^2. \label{function1}
\end{equation}
The coordinate ranges are
\begin{equation}
     -\infty < t < +\infty, \quad r \geq 0, \quad 0 \leq \theta \leq \pi, \quad 0 \leq \phi < 2\pi. \label{aa5}
\end{equation}
Throughout the discussion we consider the state parameter $w=-2/3$, so that the metric function $f(r)$ reduces as
\begin{equation}
     f(r) = 1 - \alpha - \frac{2 M}{r} - \frac{\mathcal{N}}{r^{m}} - \frac{\Lambda}{3} r^2,\quad\quad m=\frac{1+2\,\lambda}{\lambda-1}. \label{function}
\end{equation}

It is worth noting that in the limit $\lambda \to 0$, corresponding to the absence of Rastall gravity effects, one recovers the Letelier AdS BH surrounded by quintessence-like fluid in general relativity, as discussed in \cite{MMDC}.

\begin{figure}[ht!]
    \centering
    \includegraphics[width=0.45\linewidth]{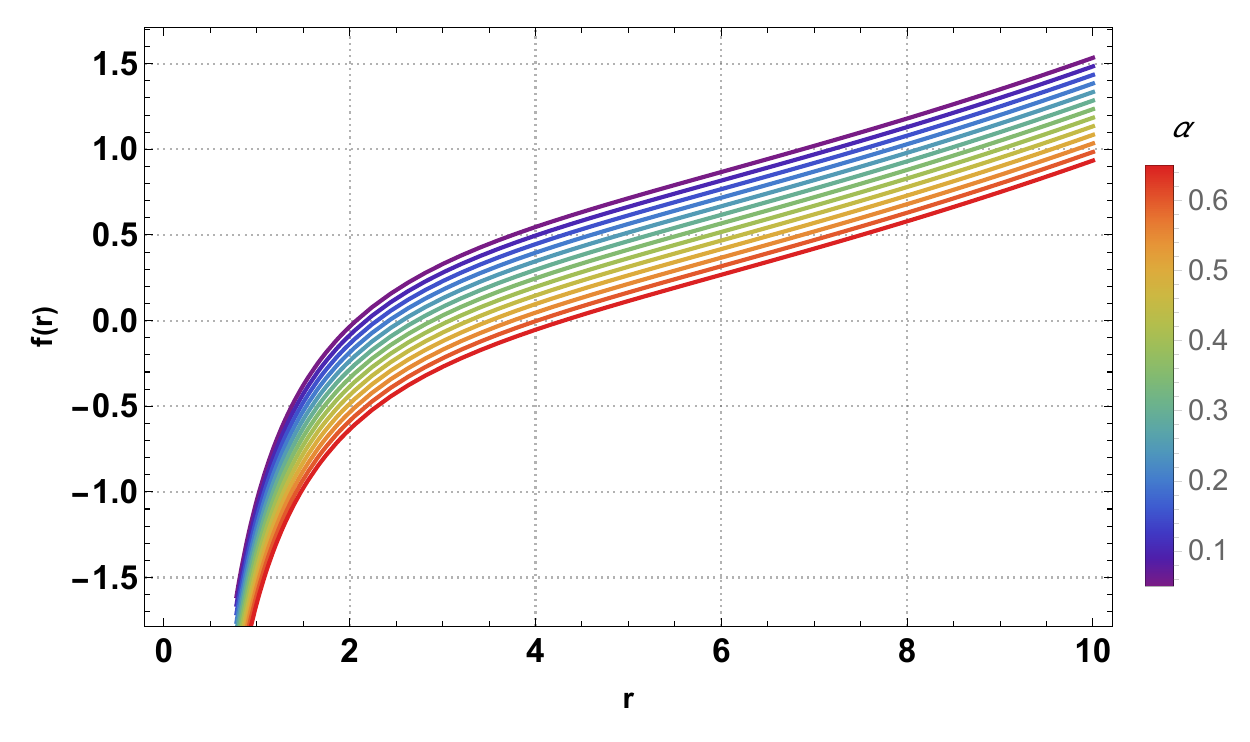}\qquad
    \includegraphics[width=0.45\linewidth]{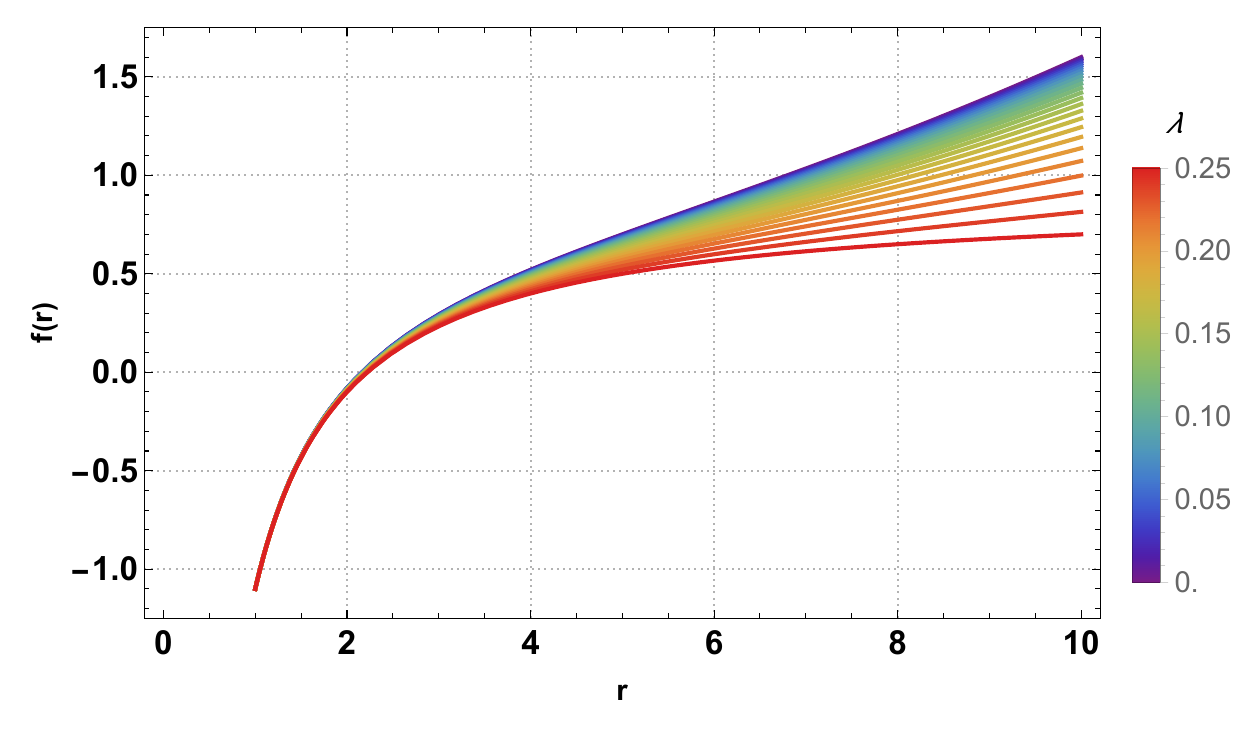}\\
    (a) $\lambda=0.1$ \hspace{8cm} (b) $\alpha=0.1$
    \caption{\footnotesize Behavior of the metric function $f$ as a function of $r$ for different values of Cos parameter $\alpha$ and the Rastall parameter $\lambda$. Here $M=1,\,\Lambda=-0.03,\,\mathcal{N}=0.01,\,w=-2/3$.}
    \label{fig:metric}
\end{figure}

In Figure \ref{fig:metric}, we illustrate the metric function $f(r)$ by varying the Cos parameter $\alpha$ and the Rastall parameter $\lambda$, while keeping other parameters fixed for a particular state parameter $w=-2/3$.

\section{Geodesic Structure}\label{sec:3}

Geodesic motion describes the trajectories of free-falling particles and light in curved spacetime around black holes, governed by the geodesic equation derived from the spacetime metric. Timelike geodesics correspond to massive particles, such as stars or gas, while null geodesics describe photon paths. Around BHs, such as Schwarzschild or Kerr solutions, geodesics exhibit unique properties including the existence of an innermost stable circular orbit (ISCO) for massive particles and a photon sphere where light can orbit in unstable circular paths. The effective potential method is commonly employed to analyze these orbits, revealing key features like orbital stability and perihelion precession caused by the strong gravitational field. Understanding geodesic motion is fundamental for interpreting phenomena such as accretion disk behavior, gravitational lensing, and the formation of BH shadows. These aspects have been extensively studied in classical works  \cite{SC,FZ,BC}.

\subsection{Null Geodesic Motions}\label{sec:3-1}

Null geodesics describe the paths of massless particles such as photons and satisfy the condition \(ds^2 = 0\). The geodesic equations can be derived via the Euler-Lagrange method or Hamilton-Jacobi formalism. These yield conserved quantities like energy $\mathrm{E}$ and angular momentum $\mathrm{L}$, reducing the problem to an effective potential analysis \cite{SC}. The effective potential determines photon trajectories and shows the photon sphere, which consists of unstable circular orbits critical for phenomena such as black hole shadows \cite{VP}. Studying null geodesics is essential for understanding gravitational lensing, light deflection, and the observational signatures of black holes, providing stringent tests for General Relativity and its alternatives \cite{JLS,EHT1}.

We start with the Lagrangian  
\begin{equation}
    \mathcal{L}=\frac{1}{2}\,g_{\mu\nu}\,\left(\frac{dx^{\mu}}{d\lambda}\right)\,\left(\frac{dx^{\nu}}{d\lambda}\right),\label{null1}
\end{equation}
where $\lambda$ is an affine parameter.

The spacetime (\ref{metric}) is spherically symmetric, thus the Lagrangian density function  becomes ($\theta=\pi/2$ )
\begin{equation}
    \mathcal{L}=\frac{1}{2}\,\left[-f(r)\,\left(\frac{dt}{d\lambda}\right)^2+\frac{1}{f(r)}\,\left(\frac{dr}{d\lambda}\right)^2+r^2\,\left(\frac{d\phi}{d\lambda}\right)^2\right].\label{null2}
\end{equation}
We have two conserved quantities corresponding to cyclic coordinates ($t, \phi$)  the conserved energy $\mathrm{E}$ and the conserved angular momentum $\mathrm{L}$:
\begin{equation}
    \mathrm{E}=f(r)\,\left(\frac{dt}{d\lambda}\right)\qquad,\qquad \mathrm{L}=r^2\,\left(\frac{d\phi}{d\lambda}\right).\label{null3}
\end{equation}

For null geodesics, $ds^2=0$, and using the metric (\ref{metric}), we have
\begin{equation}
    -f(r)\,\left(\frac{dt}{d\lambda}\right)^2 + \frac{1}{f(r)}\,\left(\frac{dr}{d\lambda}\right)^2 + r^2\,\left(\frac{d\phi}{d\lambda}\right)^2=0,\label{null4}
\end{equation}
where dot represents ordinary derivative w. r. to an affine parameter $\lambda$ along the null geodesic.

Eliminating $\dot{t}$ and $\dot{\phi}$ in Eq. (\ref{null2}) and considering the geodesic motions in the equatorial plane defined by $\theta=\pi/2$, we find
\begin{equation}
    \left(\frac{dr}{d\lambda}\right)^2+V_\text{eff}(r)=\mathrm{E}^2,\label{null5}
\end{equation}
where
\begin{equation}
    V_\text{eff}(r)=\frac{\mathrm{L}^2}{r^2}\,f(r).\label{null6}
\end{equation}

Equation (\ref{null5}) is the one-dimensional equation of motion of a particle of unit mass having energy $\mathrm{E}^2$ in a modified potential $V_\text{eff}(r)$.

We observe that the effective potential governs the dynamics of photon particles in modified by CoS parameter $\alpha$, the normalization $\mathcal{N}$ of the QF, the Rastall parameter $\lambda$. Moreover, the BH mass $M$ and the cosmological constant (CC) alter this potential. 

\begin{figure}[ht!]
    \centering
    \includegraphics[width=0.45\linewidth]{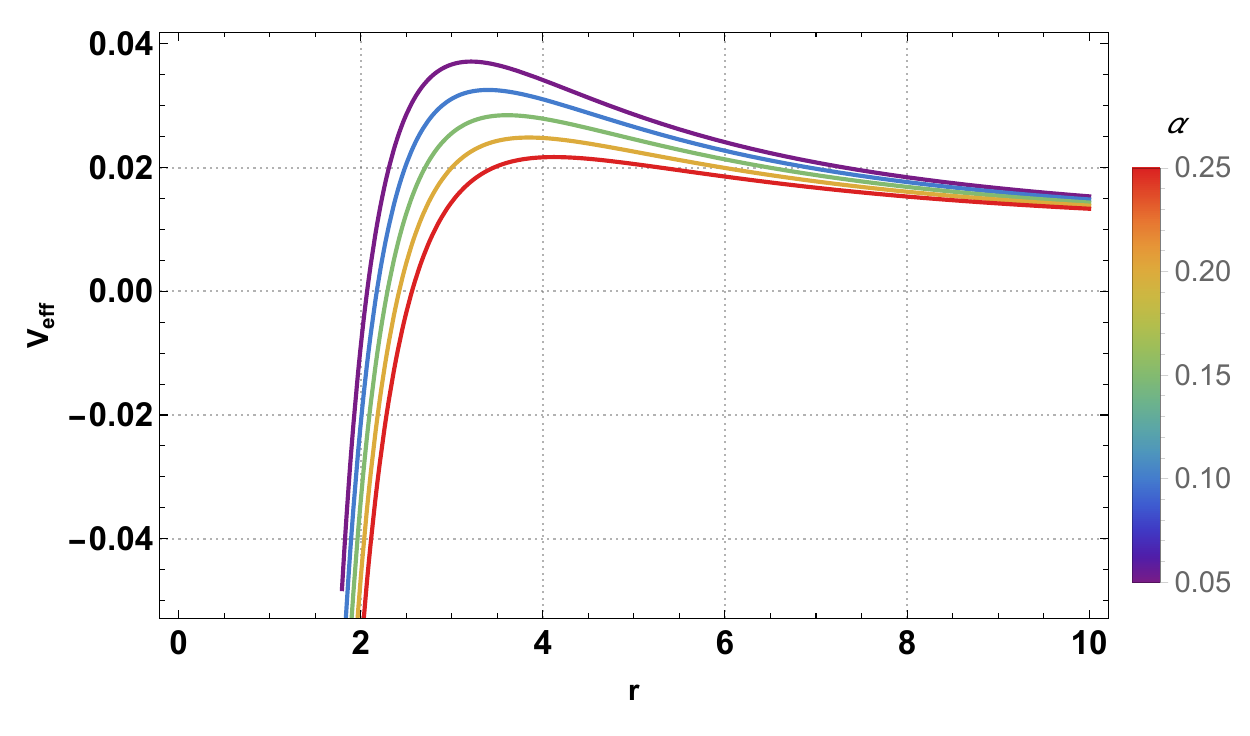}\qquad
    \includegraphics[width=0.45\linewidth]{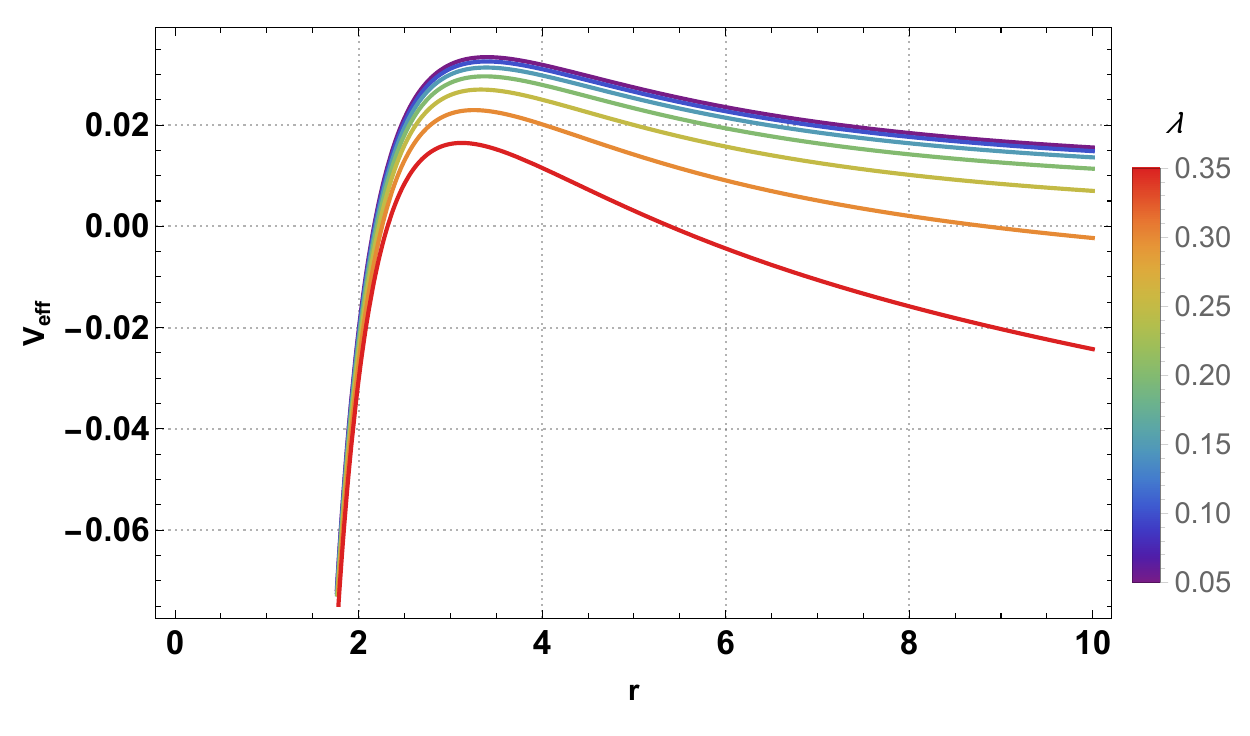}\\ 
    (a) $\lambda=0.1$ \hspace{8cm} (b) $\alpha=0.1$
    \caption{\footnotesize Behavior of the effective potential governing the photon dynamics as a function of $r$ for various values of CoS parameter $\alpha$ and Rastall parameter $\lambda$. Here, $M=1,\,\mathrm{L}=1,\,\mathcal{N}=0.01,\,\Lambda=-0.03$.}
    \label{fig:null-potential}
\end{figure}

In Figure \ref{fig:null-potential}, we depict the effective potential for null geodesics showing the effects of CoS parameter $\alpha$ and Rastall parameter $\lambda$, while keeping the QF parameters fixed. We observe that this potential decreases with increasing values of $\alpha$ in panel (a). In contrast this potential reduce with increasing $\lambda$ in panel (b).

\begin{center}
    \large{\bf I.\,Effective Radial Force}\label{sec:3-1-3}
\end{center}

The effective radial force is the force experiences by photon particles in the gravitational field. This force can be determined  as the negative gradient of the effective potential for null geodesics and is defined by
\begin{equation}
    F_{\text{eff}} = -\frac{dV_{\text{eff}}}{dr}.\label{null16}
\end{equation}

Using the potential given in Eq.~(\ref{null6}), we find the following expression
\begin{equation}
    F_{\text{eff}}=\frac{\mathrm{L}^2}{r^3}\,\left(1 - \alpha-\frac{3\,M}{r} - \frac{(2 + m)\,\mathcal{N}}{2\,r^m} \right).\label{null17}
\end{equation}

\begin{figure}[ht!]
    \centering
    \includegraphics[width=0.45\linewidth]{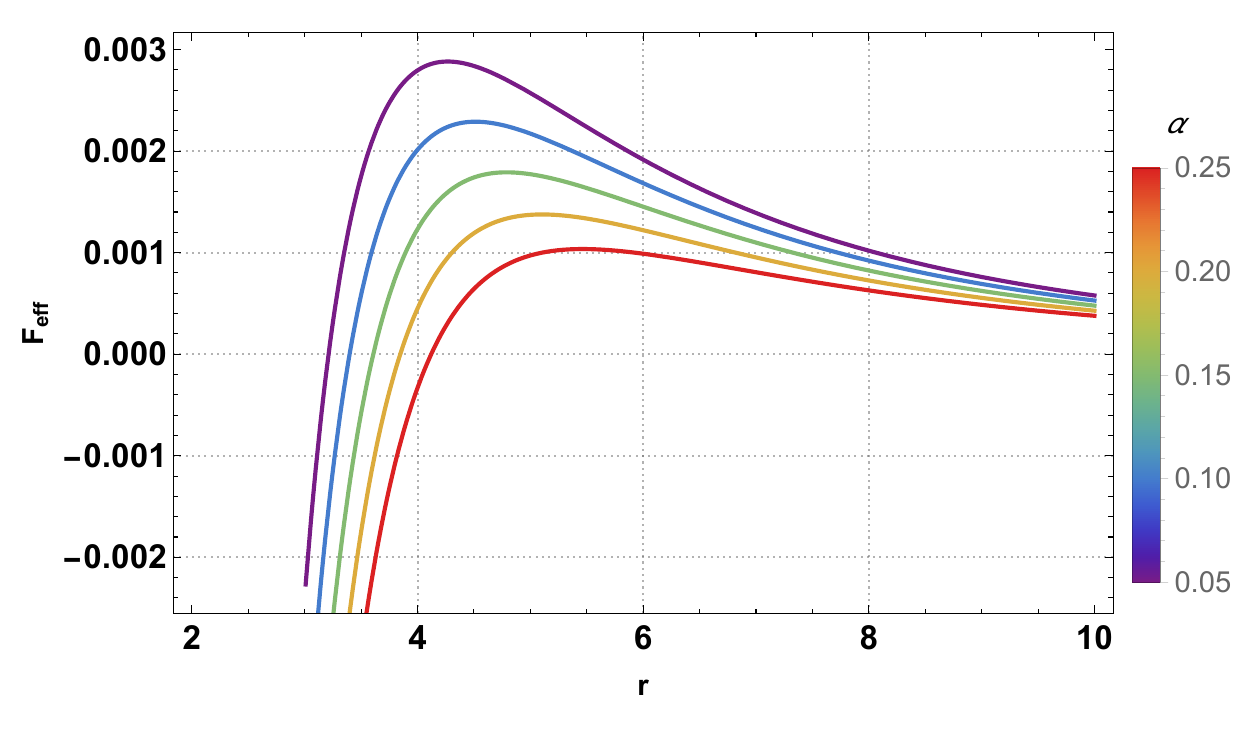}\qquad
    \includegraphics[width=0.45\linewidth]{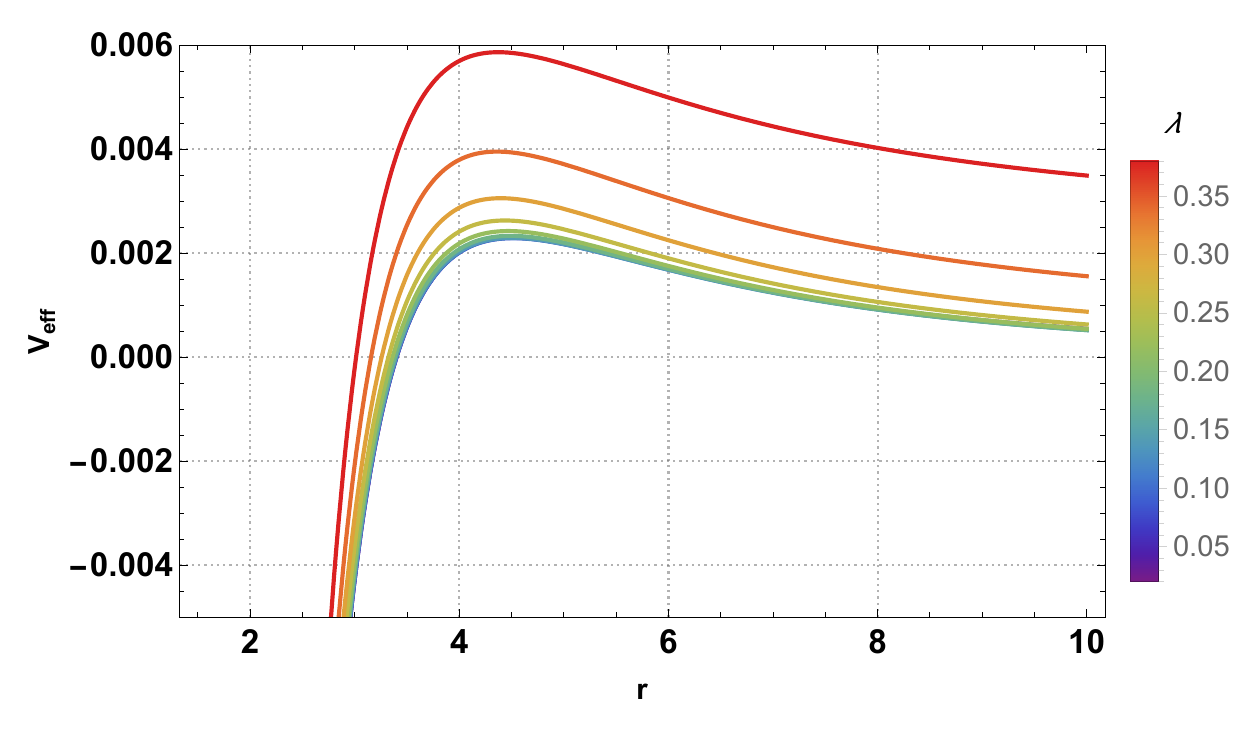}\\
     (a) $\lambda=0.1$ \hspace{8cm} (b) $\alpha=0.1$
    \caption{\footnotesize Behavior of the effective radial force $F_{eff}$ as a function of $r$ for various values of CoS parameter $\alpha$ and Rastall parameter $\lambda$. Here, $M=1,\,\mathrm{L}=1,\,\mathcal{N}=0.01,\,\Lambda=-0.03$.}
    \label{fig:force}
\end{figure}

We observe that the effective radial force experiences by photon particles given in Eq.~(\ref{null17}) is modified by the CoS parameter $\alpha$, the normalization $\mathcal{N}$ of the QF, the Rastall parameter $\lambda$. Moreover, the BH mass $M$ alters this force. 

In Figure \ref{fig:force}, we illustrate behavior of this effective radial force as a function $r$ showing the effects of CoS parameter $\alpha$ and the Rastall parameter $\lambda$, while keeping the QF parameters fixed. We observe that this radial force decreases with increasing the values of $\alpha$ in panel (a). In contrast, this force reduces with increasing $\lambda$ in panel (b).

\begin{center}
    \large{\bf II.\, Photon sphere and BH shadow} \label{sec:3-1-2}
\end{center}

The photon sphere is a spherical region surrounding a BH where gravity is strong enough that photons (light particles) can travel in circular orbits. However, these orbits are unstable, meaning any small disturbance will cause the photon either to fall into the BH center or escape to infinity. The photon sphere plays a crucial role in gravitational lensing and in the formation of the BH shadow.

For circular null orbits of radius $r=r_c$, the conditions $\dot{r}=0$ and $\ddot{r}=0$ must be satisfied. These conditions using (\ref{null5}) and (\ref{null6}) implies the following relations:
\begin{equation}
    \mathrm{E}^2=V_\text{eff}(r)=\frac{\mathrm{L}^2}{r^2}\,f(r)\Big{|}_{r=r_c}\label{null12}
\end{equation}
which gives us the critical impact parameter for photons and is given by
\begin{equation}
    \beta_c=\frac{\mathrm{L}(\mbox{ph})}{\mathrm{E}(\mbox{ph})}=\frac{r}{\sqrt{1 - \alpha - \frac{2 M}{r} - \frac{\mathcal{N}}{r^{m}} - \frac{\Lambda}{3} r^2}}\Big{|}_{r=r_c}.\label{null13}
\end{equation}
And
\begin{equation}
    \frac{d}{dr}\,\left(\frac{f(r)}{r^2}\right)=0.\label{null0}
\end{equation}
Simplifying the above relation results the following polynomial relation:
\begin{align}
    1 - \alpha - \frac{3\,M}{r} - \frac{(2 + m)\,\mathcal{N}}{2\,r^m} = 0.
\label{null11}
\end{align}

Equation (\ref{null11}) is a polynomial relation in $r$ whose exact analytical solution will give us the photon sphere radius. But, from the above polynomial relation, we see that an analytical solution is quite a challenging task without choosing a suitable value of the Rastall parameter $\lambda$.

In Table \ref{tab:1}, we present numerical values of PS radius $r_\text{ph}$ by choosing suitable values of both Cos parameter $\alpha$ and the Rastall parameter $\lambda$. We observe that as the values of $\alpha$ gradually increases, PS radius also increases for a particular value of $\lambda$. However, as we increase $\lambda$, PS slowly reduce in size for a fixed $\alpha$.

\begin{table}[h!]
\centering
\begin{tabular}{|c|c|c|c|c|}
\hline
$\alpha (\downarrow) \backslash \lambda (\rightarrow )$ & 0.05 & 0.10 & 0.15 & 0.20 \\
\hline
0.05 & 3.212905 & 3.211278 & 3.205030 & 3.189851 \\
\hline
0.12 & 3.479629 & 3.478538 & 3.471360 & 3.451956 \\
\hline
0.15 & 3.608396 & 3.607694 & 3.600099 & 3.578405 \\
\hline
0.20 & 3.846319 & 3.846598 & 3.838319 & 3.811940 \\
\hline
0.25 & 4.119112 & 4.121011 & 4.112145 & 4.079626 \\
\hline
0.30 & 4.435426 & 4.440006 & 4.430839 & 4.390094 \\
\hline
0.35 & 4.807196 & 4.816275 & 4.807495 & 4.755430 \\
\hline
\end{tabular}
\caption{\footnotesize Numerical results for the photon sphere (PS) radius $r_\text{ph}$ by varying the values of $\alpha$ and $\lambda$. Here $M=1,\,\mathcal{N}=0.01$.}
\label{tab:1}
\end{table}

Finally, we aim to find shadow size cast by the BH and analyze how the geometric and physical parameters alter the shadow size. The BH shadow is the dark region observed against the backdrop of bright emission from accreting matter or surrounding light sources, caused by the strong gravitational lensing and photon capture by the BH. It corresponds to the apparent shape of the photon capture sphere as seen by a distant observer. The shadow is larger than the event horizon because it includes light that was bent around the photon sphere before being captured. Mathematically, the shadow boundary is determined by the unstable circular photon orbits (photon sphere) around the BH \cite{VP,TJ,EHT1}. 

For an observer located at infinity, the shadow radius $R_s$ s given by the critical impact parameter $\beta_c$:
\begin{equation}
    R_s=\beta_c=\frac{r_\text{ph}}{\sqrt{f(r_\text{ph})}}.\label{null14}
\end{equation}
Substituting the metric function $f(r)$ yields
\begin{equation}
    R_s=\frac{r_\text{ph}}{\sqrt{1 - \alpha - \frac{2 M}{r_\text{ph}} - \frac{\mathcal{N}}{r_\text{ph}^{m}} - \frac{\Lambda}{3}\,r_\text{ph}^2}}.\label{null15}
\end{equation}

We observe that the shadow size given in Eq.~(\ref{null15}) is modified by the CoS parameter $\alpha$, the normalization $\mathcal{N}$ of the QF, the Rastall parameter $\lambda$. Moreover, the BH mass $M$ alters this radius. 

\begin{table}[ht!]
\centering
\begin{tabular}{|c|c|c|c|c|}
\hline
$\alpha (\downarrow) \backslash \lambda (\rightarrow )

$ & 0.05 & 0.10 & 0.15 & 0.20 \\
\hline
0.05 & 5.13092 & 5.18934 & 5.27385 & 5.40245 \\
\hline
0.10 & 5.61569 & 5.69303 & 5.80765 & 5.98773 \\
\hline
0.15 & 5.84013 & 5.92739 & 6.05820 & 6.26692 \\
\hline
0.20 & 6.23640 & 6.34306 & 6.50626 & 6.77419 \\
\hline
0.25 & 6.65916 & 6.78923 & 6.99283 & 7.33807 \\
\hline
0.30 & 7.10510 & 7.26299 & 7.51637 & 7.96212 \\
\hline
0.35 & 7.56826 & 7.75840 & 8.07200 & 8.64721 \\
\hline
\end{tabular}
\caption{Computed values of the shadow size $R_s$ by varying $\alpha$ and $\lambda$. Here $M=1,\,\mathcal{N}=0.01,\,\Lambda=-0.03$.}
\label{tab:2}
\end{table}

In Table \ref{tab:2}, we present numerical values of shadow radius $R_s$ by choosing suitable values of both the Cos parameter $\alpha$ and the Rastall parameter $\lambda$, while keeping QF parameters fixed. In contrast to PS radius, here we observe that as the values of $\alpha$ and $\lambda$ gradually increases, shadow radius also increases. Both parameters simultaneously affect the size of the shadow cast by the selected BH.

\begin{center}
    \large{\bf III.\, Lyapunov Exponent: Criteria for Unstable Circular Null Orbits}
\end{center}

Circular null orbits, or photon spheres, represent the paths along which massless particles like photons can orbit a BH in a circular trajectory. These orbits are generally unstable, meaning that small perturbations cause photons to either fall into the black hole or escape to infinity. The stability of such orbits can be analyzed using the effective potential \( V_{\mathrm{eff}}(r) \) of null geodesics; specifically, an unstable circular null orbit corresponds to a local maximum of \( V_{\mathrm{eff}} \), where the second derivative satisfies \( V_{\mathrm{eff}}''(r_c) < 0 \), with \( r_c \) denoting the radius of the circular orbit. The instability timescale of these orbits is quantified by the Lyapunov exponent \(\lambda_L\), which measures the exponential rate at which nearby trajectories diverge. This exponent can be expressed as
\begin{equation}
\lambda_L = \sqrt{\frac{-V_{\mathrm{eff}}''(r_c)}{2\, \dot{t}^2}},\label{condition-1}
\end{equation}
where \( \dot{t} \) is the time component of the four-velocity of the photon. A larger Lyapunov exponent indicates a more rapid departure from the circular orbit, linking the orbit’s instability to observable phenomena such as the decay rates of quasinormal modes and the black hole’s ringdown signals.

Using relations Eq.~(\ref{null3}), (\ref{null6}) and (\ref{null12}), we find the following expression
\begin{equation}
    \lambda_L = \sqrt{\left(1 - \alpha - \frac{2 M}{r} - \frac{\mathcal{N}}{r^{m}} - \frac{\Lambda}{3} r^2\right)\,\left(\frac{1 - \alpha}{r^{2}} + \frac{\mathcal{N}}{r^{m+2}}\,\left(\frac{m(m+1)}{2} - 1\right)\right)}\Bigg{|}_{r=r_c}.\label{condition-2}
\end{equation}

\begin{figure}[ht!]
    \centering
    \includegraphics[width=0.45\linewidth]{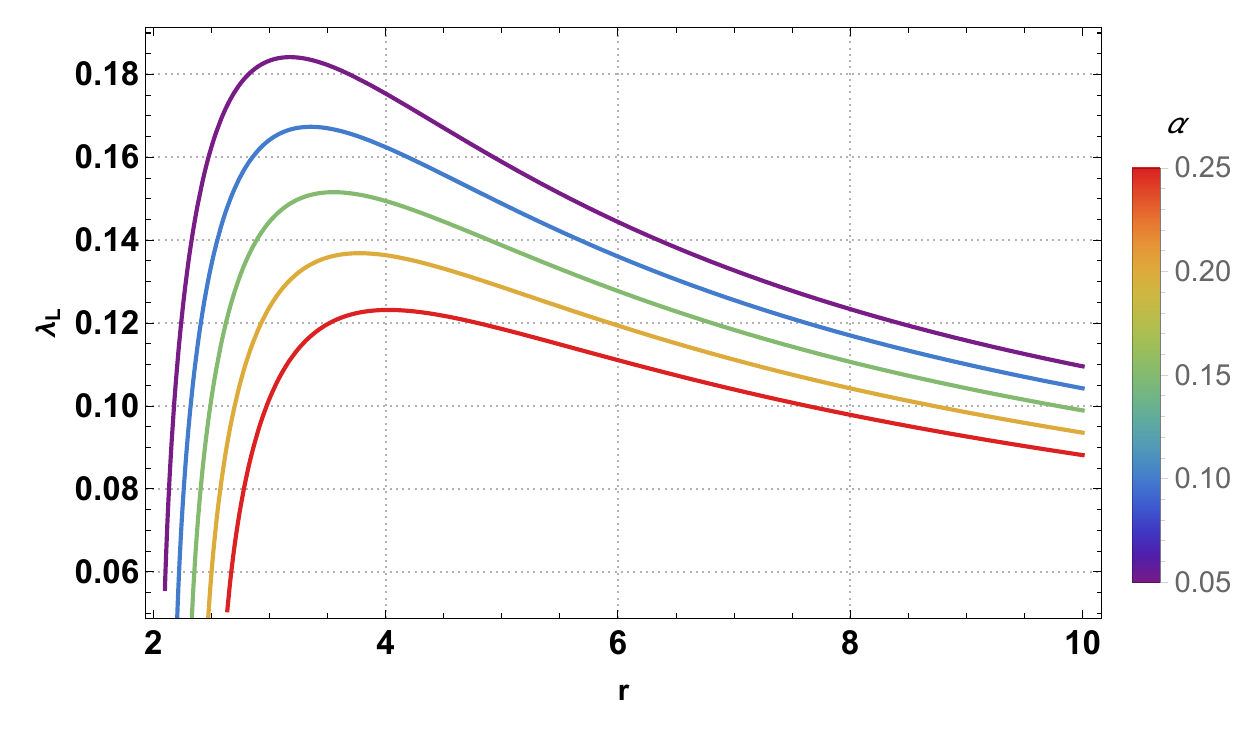}
    \caption{\footnotesize Behavior of the Lyapunov exponent  $\lambda_L$ as a function of $r=r_c$ for various values of CoS parameter $\alpha$. Here, $M=1,\,\lambda=0.1,\,\mathcal{N}=0.01,\,\Lambda=-0.03$.}
    \label{fig:lyapunov}
\end{figure}

One can observe that the stability of circular null orbits depends on the CoS parameter $\alpha$, the normalization $\mathcal{N}$ of the QF, the Rastall parameter $\lambda$. Moreover, the BH mass $M$ and CC $\Lambda$ also alters the Lyapunov exponent quantity. 

In Figure \ref{fig:lyapunov}, we show the nature of Lyapunov exponent $\lambda_L$ by varying CoS parameter $\alpha$, while keeping all other parameters fixed. From this Figure, we observe that $\lambda_L$ remains positive definite, which confirms unstable circular null orbits.

\begin{center}
    \large{\bf IV.\, Topological Features of Photon Rings}\label{sec:3-1-4}
\end{center}

Topological methods have recently become powerful tools for analyzing BH solutions. In particular, Duan’s \(\phi\)-mapping theory connects topological defects—arising where vector fields vanish-with critical points and phase transitions in BH systems. These defects generate a conserved topological current, leading to a topological invariant that encodes the system’s global phase structure through geometric properties of the vector field. For comprehensive studies and applications in various BH contexts, see~\cite{AA1,AA2,AA3,AA4,AA5,AA6,AA7,AA8,AA9}.

In order to study the topological property of the photon rings (PR), one can introduce a potential function as follows: \cite{AA1,AA2,AA3,AA5}
\begin{equation}
H(r,\theta) = \sqrt{\frac{-g_{tt}}{g_{\phi\phi}}} =\frac{\sqrt{f(r)}}{r\,\sin \theta}.\label{null18}
\end{equation}
Obviously, the radius of the PS locates at the root of $\partial_r\,H(r,\theta)=0$. Similar to \cite{AA2,AA3}, we can introduce a vector ${\bf v}=(v_r,\,v_{\theta})$:
\begin{equation}
    v_r=\frac{\partial_r H}{\sqrt{g_{rr}}}\quad,\quad v_{\theta}=\frac{\partial_{\theta} H}{\sqrt{g_{\theta\theta}}}.\label{null19}
\end{equation}
It follows that $\partial^{\mu} H\,\partial_{\mu} H=v_r^2+v^2_{\theta}=v^2$. Although the circular photon orbit for a spherically symmetric BH is a PS, which is independent of the coordinate $\theta$, here we aim to investigate the topological property of the circular photon orbit, so we preserve $\theta$ in
our discussions. Note that the vector field ${\bf v}$ can also be reformulated as
\begin{equation}
    {\bf v}=v\,e^{i\,\Omega}\quad,\quad v_r=v\,\cos \Omega,\quad v_{\theta}=v\,\sin \Omega.\label{null20}
\end{equation}
The normalized vectors are defined as
\begin{equation}
    {\bf n}=(n_r\,,\,n_{\theta})=\frac{{\bf v}}{v}=\left(\frac{v_r}{\sqrt{v^2_r+v^2_{\theta}}}\,,\,\frac{v_{\theta}}{\sqrt{v^2_r+v^2_{\theta}}}\right).\label{null21}
\end{equation}

\begin{figure}[ht!]
    \centering
    \includegraphics[width=0.3\linewidth]{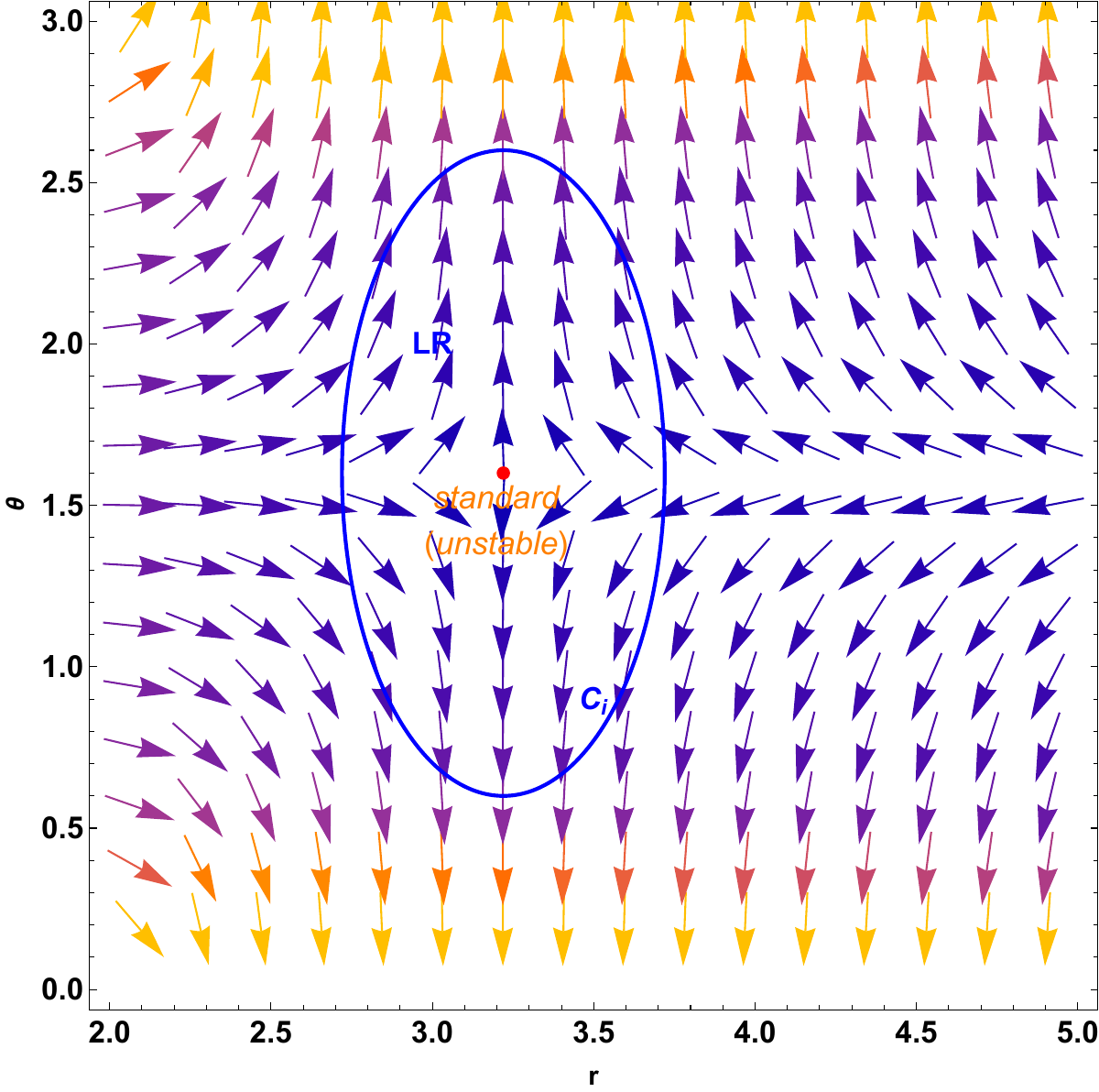}\qquad
    \includegraphics[width=0.3\linewidth]{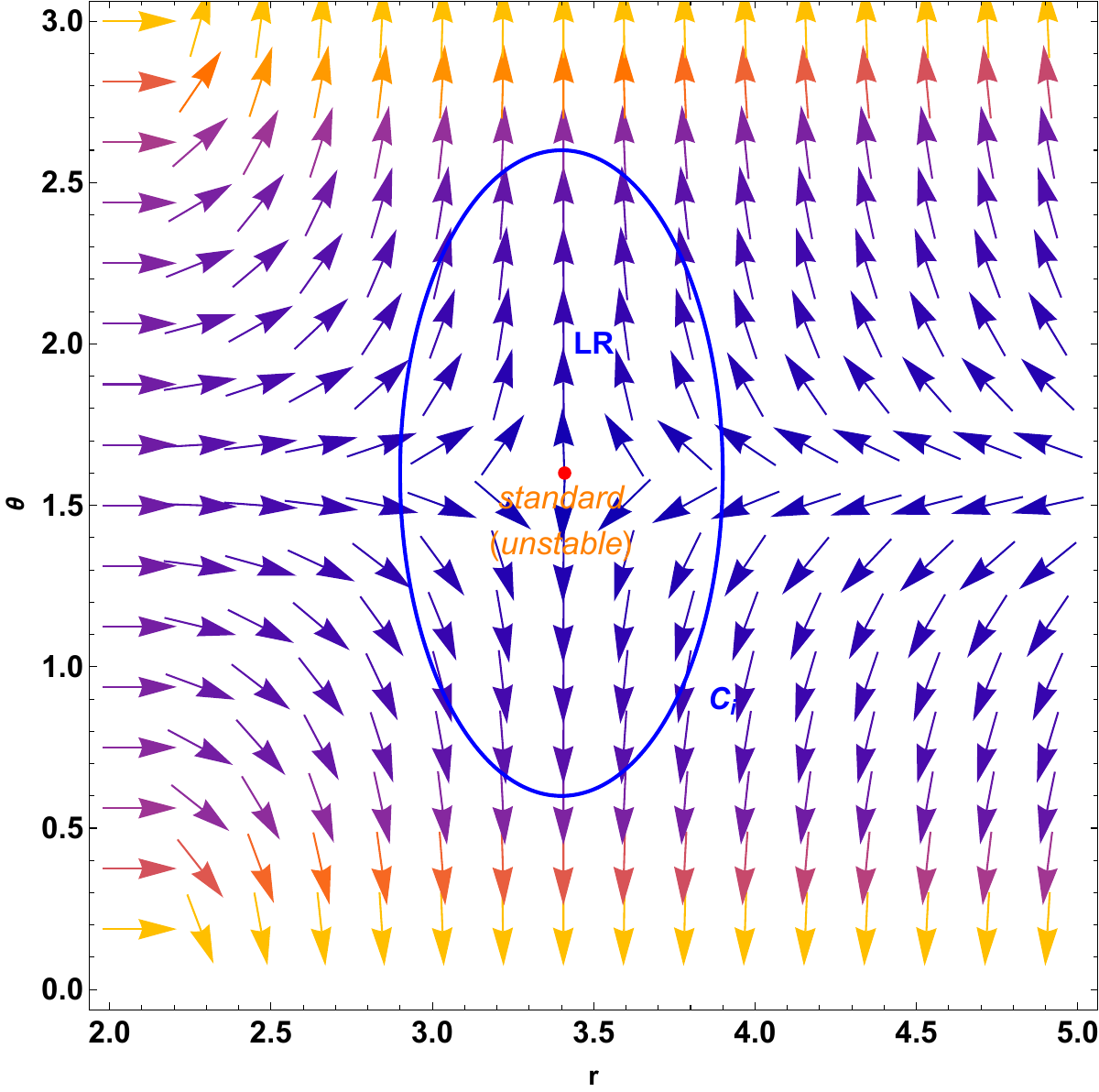}\qquad
    \includegraphics[width=0.3\linewidth]{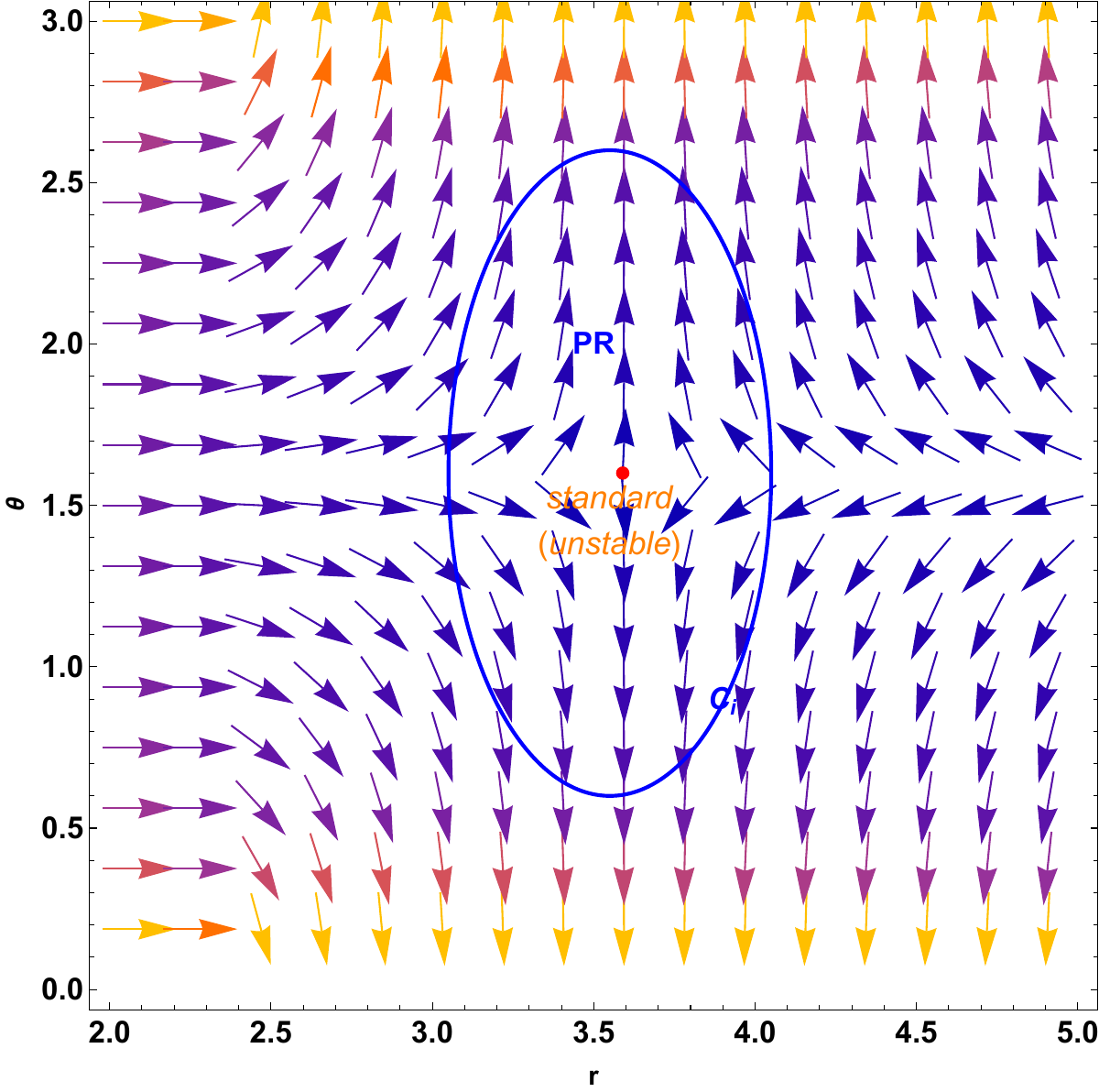}\\
     (a) $\alpha=0.05$ \hspace{4cm} (b) $\alpha=0.1$ \hspace{4cm} (c) $\alpha=0.15$
    \caption{\footnotesize The arrows represent the vector field $\mathbf{v}$ on a portion of the $r{-}\Theta$ plane for the BH. The photon ring (PR), marked with a red dot, is at $(r, \theta) = (3.22,\, \pi/2); (3.41,\,\pi/2);(3.59,\,\pi/2)$. The blue contour $\mathcal{C}_i$ is a closed loop enclosing the light ring. Here, $M=1,\lambda=0.1,\,\,\mathcal{N}=0.01,\,\Lambda=-0.03$.}
    \label{fig:topology-1}
\end{figure}

\begin{figure}[ht!]
    \centering
    \includegraphics[width=0.3\linewidth]{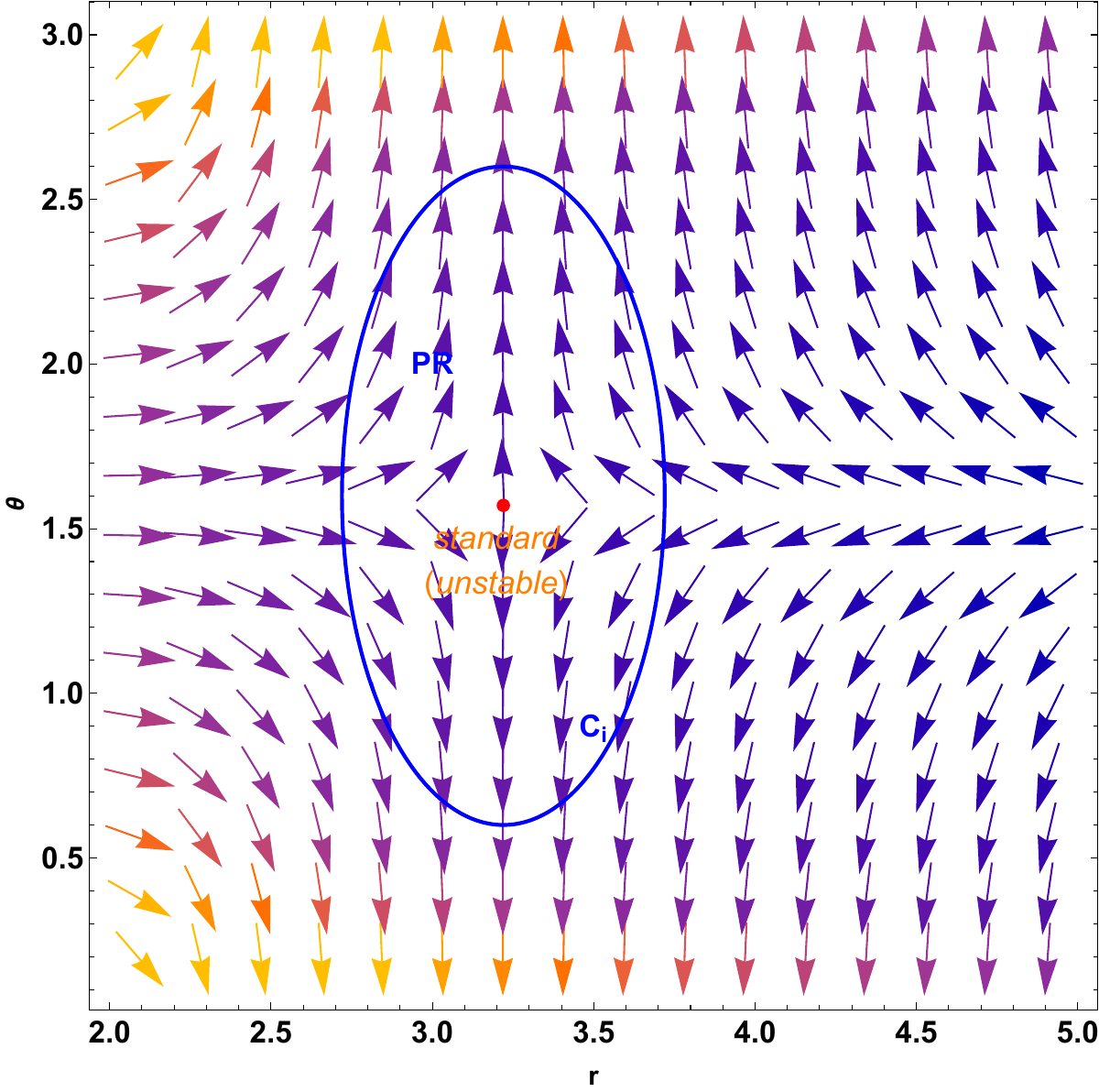}\qquad
    \includegraphics[width=0.3\linewidth]{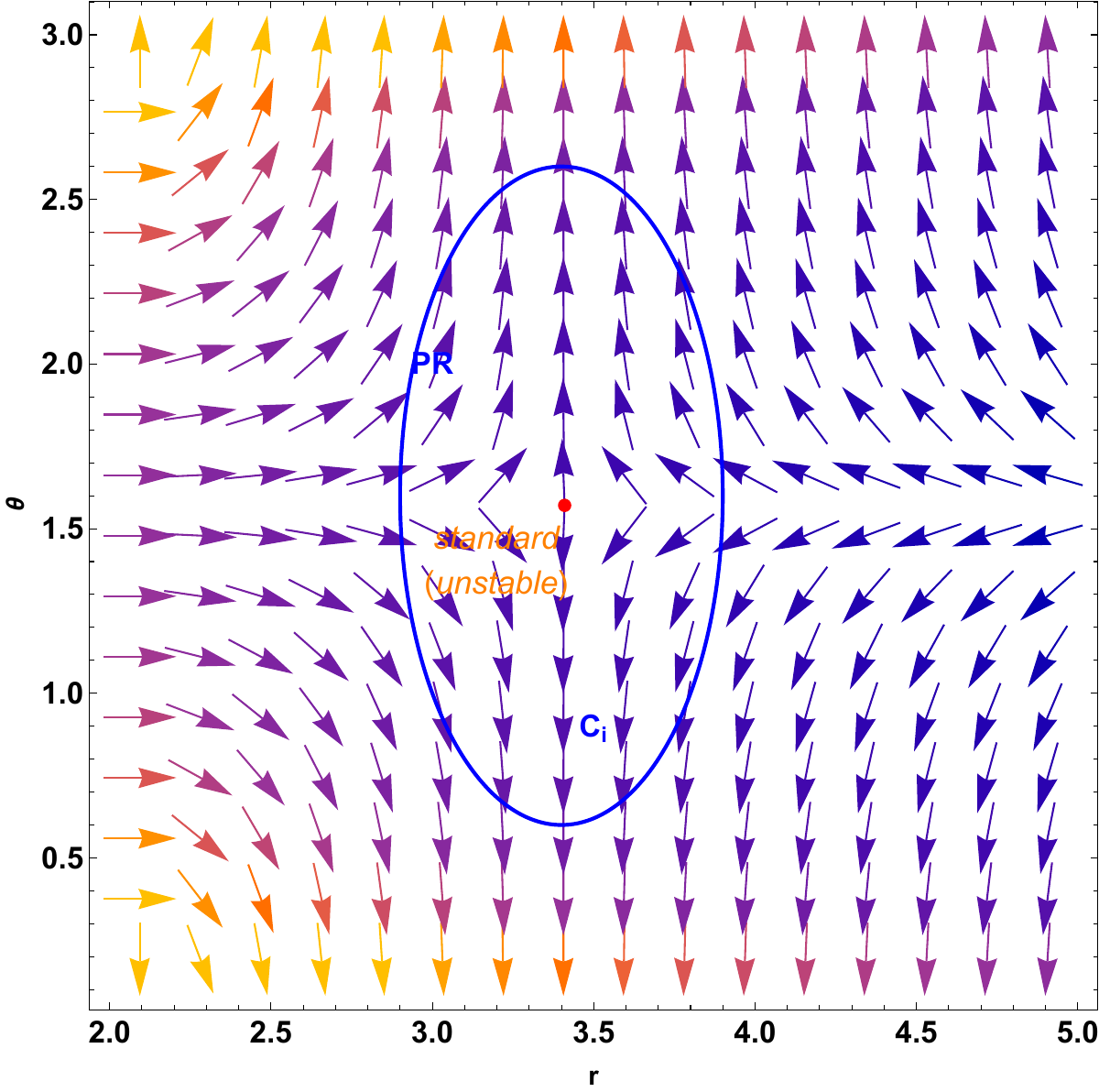}\qquad
    \includegraphics[width=0.3\linewidth]{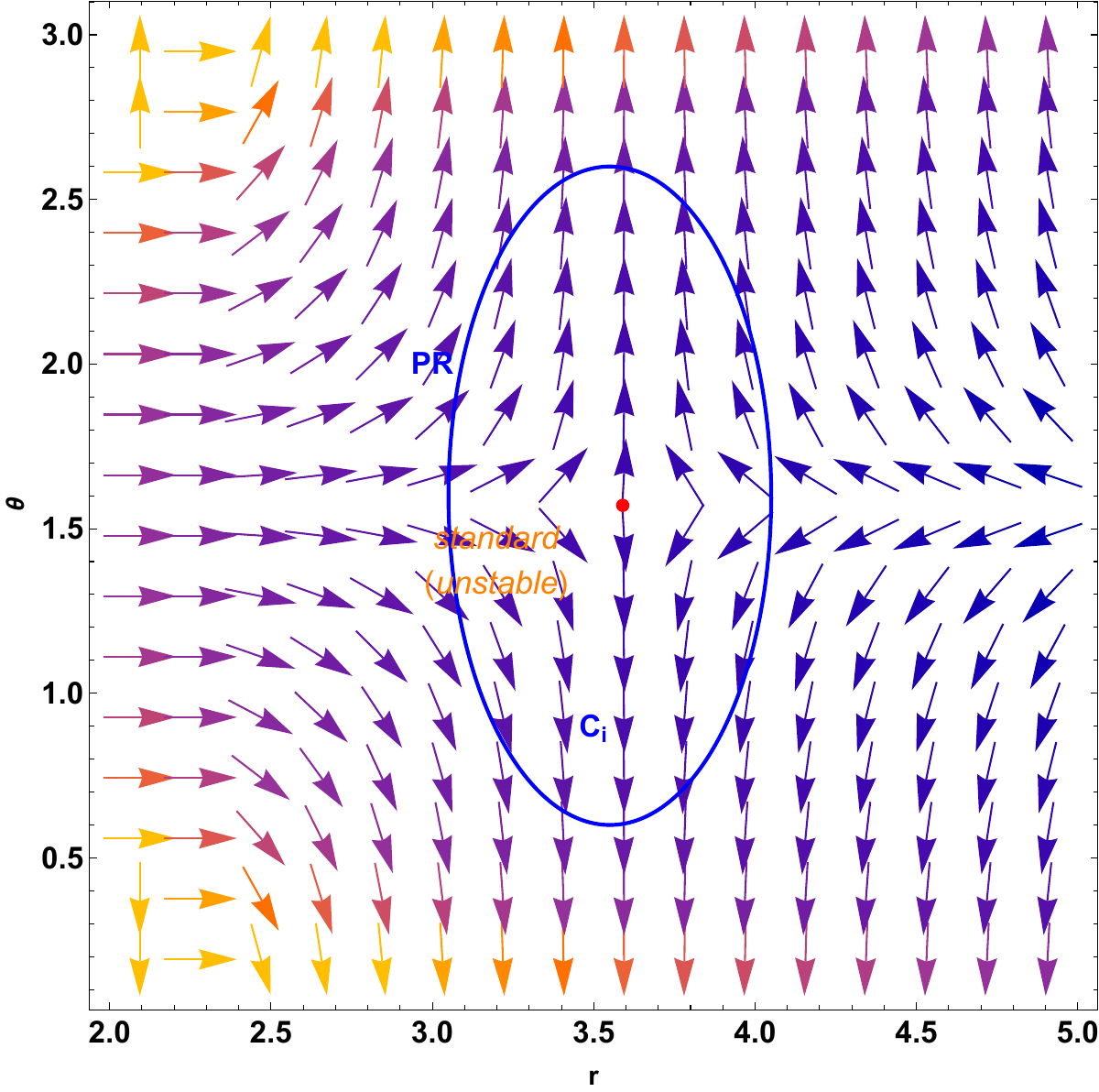}\\
     (a) $\alpha=0.05$ \hspace{4cm} (b) $\alpha=0.1$ \hspace{4cm} (c) $\alpha=0.15$
    \caption{\footnotesize The arrows represent the unit vector field $\mathbf{n}$ on a portion of the $r{-}\Theta$ plane for the BH. The photon ring (PR), marked with a red dot, is at $(r, \theta) = (3.22,\, \pi/2); (3.41,\,\pi/2);(3.59,\,\pi/2)$. The blue contour $\mathcal{C}_i$ is a closed loop enclosing the light ring. Here, $M=1,\lambda=0.1,\,\,\mathcal{N}=0.01,\,\Lambda=-0.03$.}
    \label{fig:topology-2}
\end{figure}

In our case, we find the vector field components $v_r$ and $v_{\theta}$ as,
\begin{align}
    v_r&=-\frac{1}{r^2\,\sin  \theta}\,\left(1-\alpha-\frac{3\,M}{r}-\frac{(2 + m)\,\mathcal{N}}{2\,r^m}\right)\Bigg{|}_{r_0,\theta_0},\label{null25}\\
    v_{\theta}&=-\frac{\cot \theta}{r^2\,\sin \theta}\,\sqrt{1 - \alpha - \frac{2 M}{r} - \frac{\mathcal{N}}{r^{m}} - \frac{\Lambda}{3} r^2}\Bigg{|}_{r_0,\theta_0}.\label{null26}
\end{align}
Here $(r_0, \theta_0)$ represents the standard point at which vector field vanishes. Noted that at $r_0=r_\text{ph}$ and $\theta_0=\pi/2$ both the components vanish.

The magnitude of the vector field is given by
\begin{equation}
    v=\frac{1}{r^2\,\sin\theta}\,\sqrt{\cot^2 \theta\,\left(1 - \alpha - \frac{2 M}{r} - \frac{\mathcal{N}}{r^{m}} - \frac{\Lambda}{3} r^2\right)+\left(1-\alpha-\frac{3\,M}{r}-\frac{(2 + m)\,\mathcal{N}}{2\,r^m}\right)^2}\Bigg{|}_{r_0,\theta_0}.\label{null27}
\end{equation}

Thereby, the normalized vector field components 
\begin{align}
    n_r&=-\frac{\left(1-\alpha-\frac{3\,M}{r}-\frac{(2 + m)\,\mathcal{N}}{2\,r^m}\right)}{\sqrt{\cot^2 \theta\,\left(1 - \alpha - \frac{2 M}{r} - \frac{\mathcal{N}}{r^{m}} - \frac{\Lambda}{3} r^2\right)+\left(1-\alpha-\frac{3\,M}{r}-\frac{(2 + m)\,\mathcal{N}}{2\,r^m}\right)^2}}\Bigg{|}_{r_0,\theta_0},\label{null28}\\
    n_{\theta}&=-\cot \theta\,\frac{\sqrt{1 - \alpha - \frac{2 M}{r} - \frac{\mathcal{N}}{r^{m}} - \frac{\Lambda}{3} r^2}}{\sqrt{\cot^2 \theta\,\left(1 - \alpha - \frac{2 M}{r} - \frac{\mathcal{N}}{r^{m}} - \frac{\Lambda}{3} r^2\right)+\left(1-\alpha-\frac{3\,M}{r}-\frac{(2 + m)\,\mathcal{N}}{2\,r^m}\right)^2}}\Bigg{|}_{r_0,\theta_0}.\label{null29}
\end{align}

From the above expressions (\ref{null28})--(\ref{null29}), We observe that the normalized vector is modified by the CoS parameter $\alpha$, the normalization $\mathcal{N}$ of the QF, the Rastall parameter $\lambda$. Moreover, the BH mass $M$ and CC $\lambda$ alters this. 

Figures \ref{fig:topology-1}--\ref{fig:topology-2} illustrates the behavior of the vector field ${\bf v}(r, \theta)$ and normalized vector field ${\bf n}(r, \theta)$ on a portion of the $r$-$\theta$ plane for different values of the CoS parameter $\alpha$, with all other parameters kept fixed.

\subsection{Timelike Geodesics and ISCOs}\label{sec:3-2}

Time-like geodesics describe the motion of massive particles under the influence of gravity alone, without any non-gravitational forces acting on them. These geodesics represent the natural trajectories of particles moving through curved spacetime, especially around compact astrophysical objects such as BHs. Studying time-like geodesics provides crucial insights into phenomena like accretion disk dynamics, stellar orbits near supermassive BHs, and the structure of spacetime near event horizons. These trajectories have been extensively analyzed in various BH spacetimes, including Schwarzschild, Reissner–Nordström, and Kerr geometries \cite{SC, BC, RMW}. Key aspects such as the innermost stable circular orbit (ISCO), periastron precession, and orbital stability are fundamental to both theoretical studies and astrophysical observations \cite{VPF}.

For timelike geodesic, using the conditions $g_{\mu\nu}\,\dot{x}^{\mu}\,\dot{x}^{\nu}=-1$, we find the equations of motion
\begin{align}
    &\dot{t}=\frac{\mathrm{E}_0}{f(r)},\label{timelike-1}\\
    &\dot{r}^2+V_\text{eff}(r)=\mathrm{E}^2_0,\label{timelike-2}\\
    &\dot{\phi}=\frac{\mathrm{L}_0}{r^2}.\label{timelike-3}
\end{align}
Here $\mathrm{E}_0$ and $\mathrm{L}_0$ are the conserved energy and angular momentum of massive test particles and the effective potential $V_\text{eff}(r)$ is given by
\begin{equation}
    V_\text{eff}(r)=f(r)\,\left(1+\frac{\mathrm{L}^2_0}{r^2}\right).\label{timelike-4}
\end{equation}

\begin{figure}[ht!]
    \centering
    \includegraphics[width=0.45\linewidth]{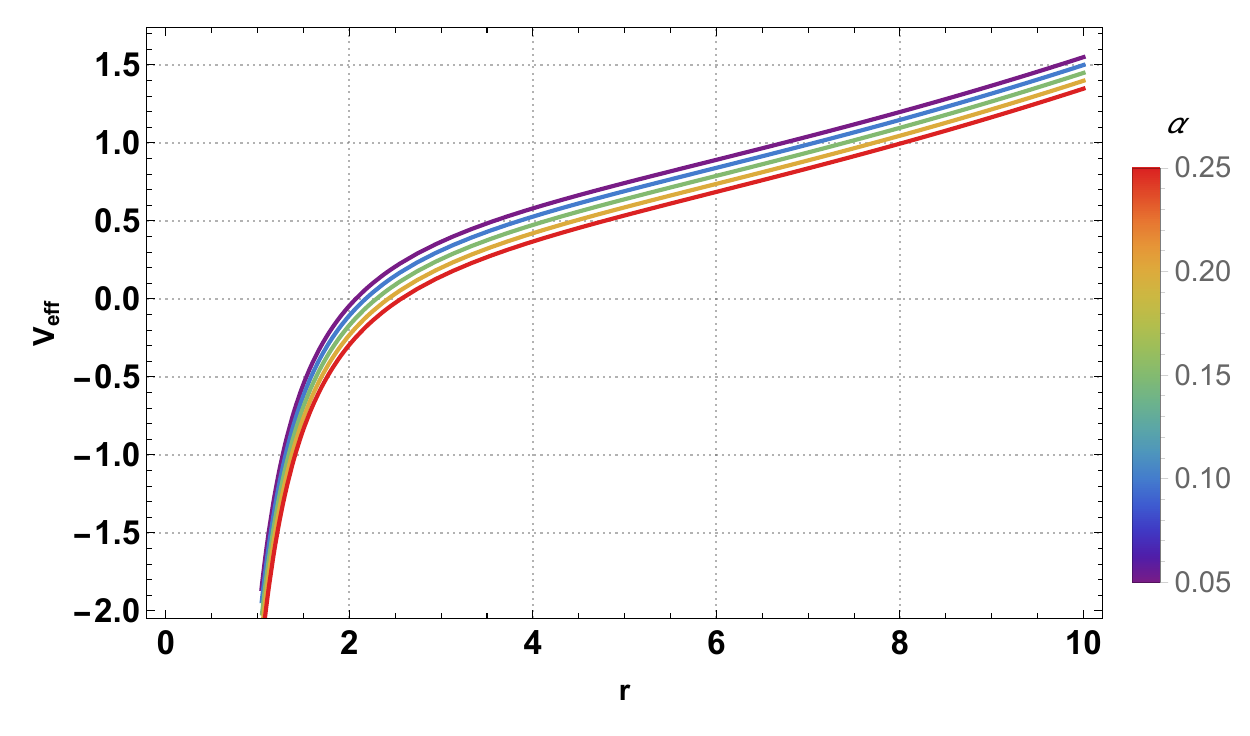}\qquad
    \includegraphics[width=0.45\linewidth]{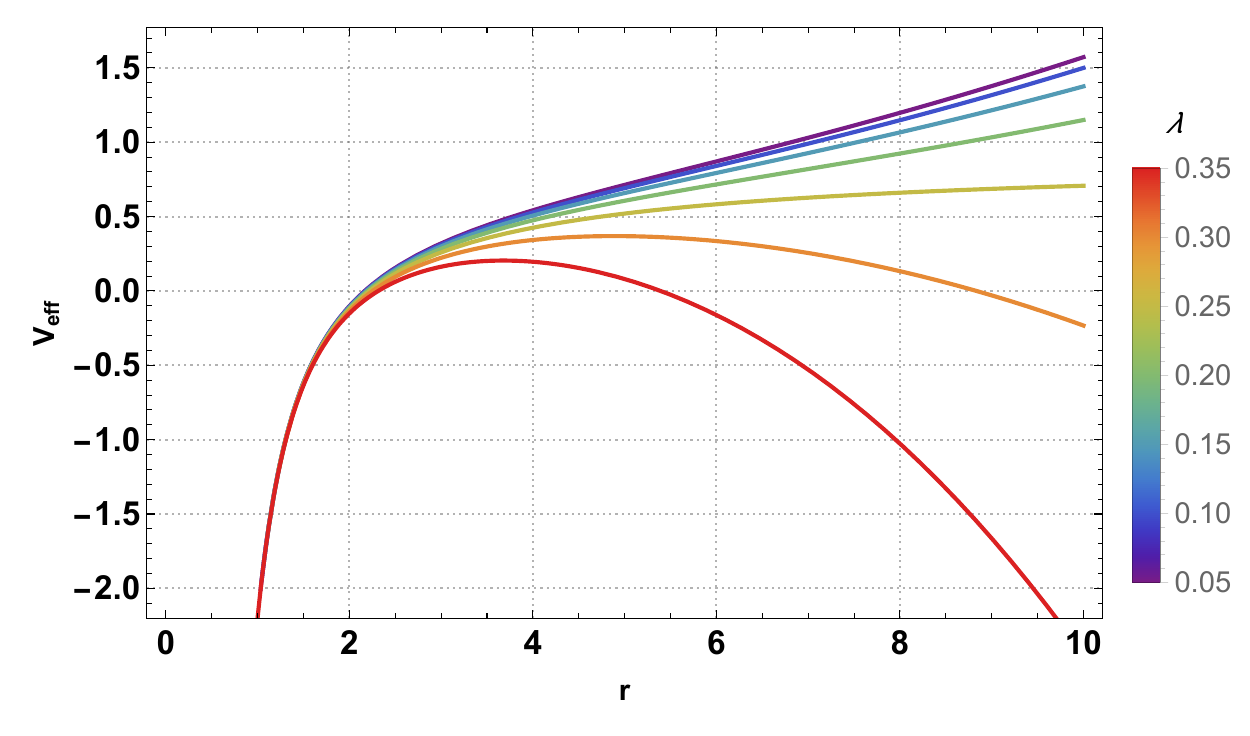}\\
     (a) $\lambda=0.1$ \hspace{8cm} (b) $\alpha=0.1$
    \caption{\footnotesize Behavior of the effective potential governing the dynamics of massive test particles as a function of $r$ for various values of CoS parameter $\alpha$ and Rastall parameter $\lambda$. Here, $M=1,\,\mathrm{L}=1,\,\mathcal{N}=0.01,\,\Lambda=-0.03$.}
    \label{fig:timelike-potential}
\end{figure}

In Figure \ref{fig:timelike-potential}, we depict the effective potential for time-like geodesics showing the effects of CoS parameter $\alpha$ and Rastall parameter $\lambda$, while keeping the QF parameters fixed. In both panels, we observe that this potential gradually decreases with increasing values of $\alpha$ and $\lambda$.

For innermost stable circular (ISCO), we have the following conditions
\begin{align}
    &\mathrm{E}^2_0=f(r)\,\left(1+\frac{\mathrm{L}^2_0}{r^2}\right),\label{timelike-5}\\
    &V'_\text{eff}(r)=0,\label{timelike-6}\\
    &V''_\text{eff}(r) \geq 0.\label{timelike-7}
\end{align}

Simplification of the condition (\ref{timelike-6}) using (\ref{timelike-4}) and with the help of \ref{timelike-5} results
\begin{equation}
    \mathrm{L}_0(\mbox{specific})=r\,\left(\frac{\frac{M}{r}+\frac{m\,\mathcal{N}}{2\,r^m}-\frac{\Lambda}{3}\,r^2}{1 - \alpha- \frac{3\,M}{r} - \frac{(2 + m)\mathcal{N}}{2\,r^m}}\right)^{1/2}.\label{timelike-8}
\end{equation}
Substituting $\mathrm{L}_0$ into the Eq. (\ref{timelike-5}) results
\begin{equation}
    \mathrm{E}_0(\mbox{specific})=\pm\,\frac{\left(1-\alpha-\frac{2\,M}{r}-\frac{\mathcal{N}}{r^m}-\frac{\Lambda}{3}\,r^2\right)}{\sqrt{1 - \alpha- \frac{3\,M}{r} - \frac{(2 + m)\mathcal{N}}{2\,r^m}}}.\label{timelike-9}
\end{equation}
Here, $\mathrm{L}_0(\mbox{specific})$ and $\mathrm{E}_0(\mbox{specific})$, respectively are the specific angular momentum and specific energy of massive test particles orbiting in circular geodesics. 

\begin{figure}[ht!]
    \centering
    \includegraphics[width=0.45\linewidth]{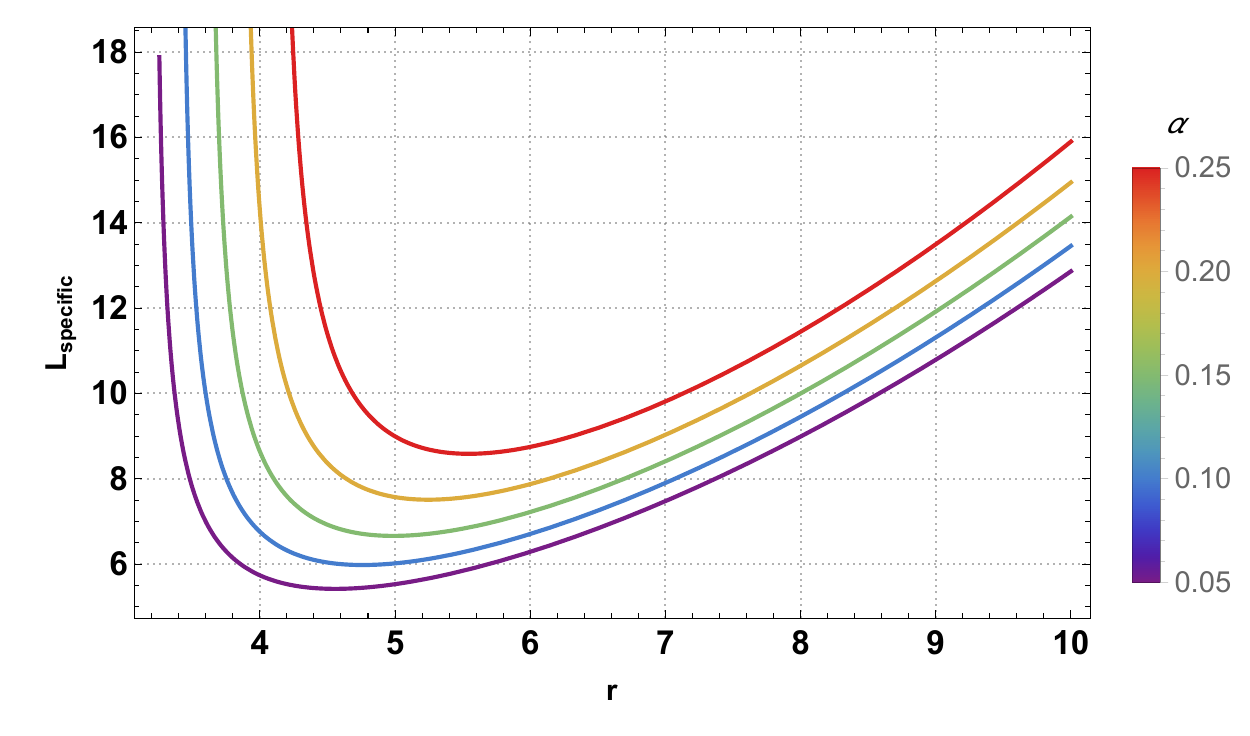}\qquad
    \includegraphics[width=0.45\linewidth]{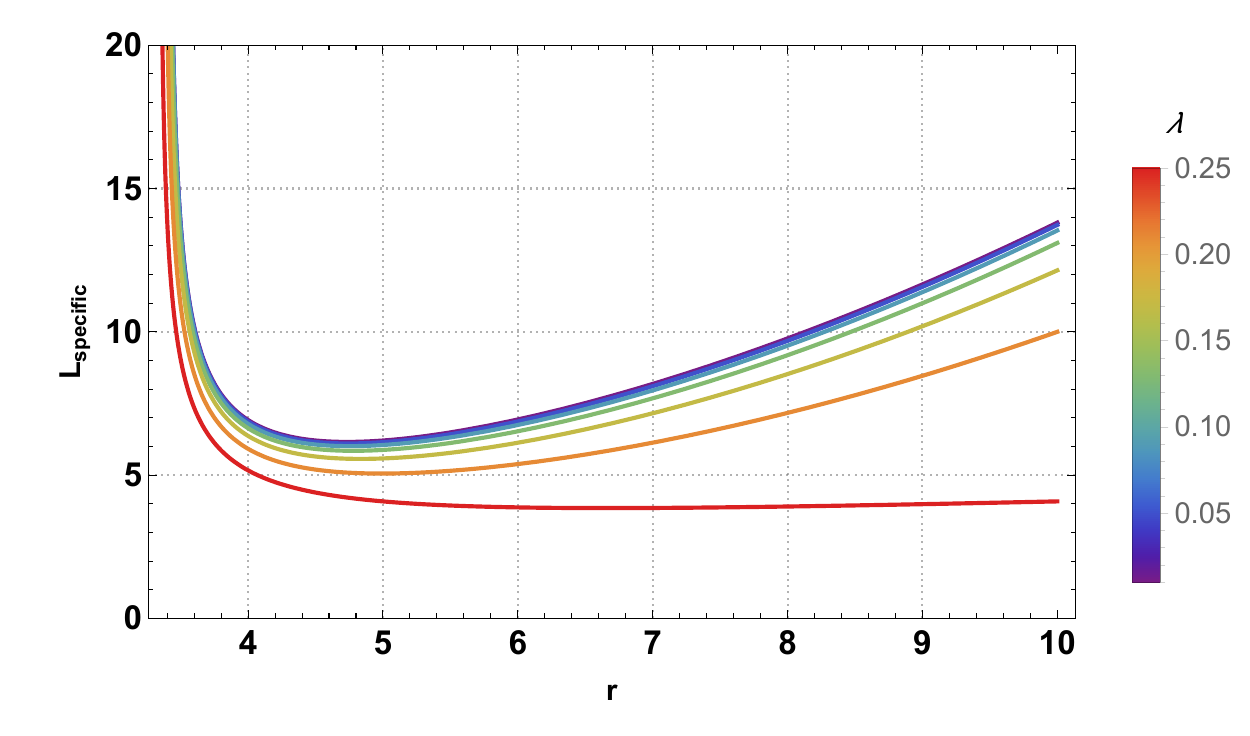}\\
     (a) $\lambda=0.1$ \hspace{8cm} (b) $\alpha=0.1$
    \caption{\footnotesize Behavior of the specific angular momentum $\mathrm{L}_0(\mbox{specific})$ as a function of $r$ for various values of CoS parameter $\alpha$ and Rastall parameter $\lambda$. Here, $M=1,\,\mathrm{L}=1,\,\mathcal{N}=0.01,\,\Lambda=-0.03$.}
    \label{fig:angular}
    \centering
    \includegraphics[width=0.45\linewidth]{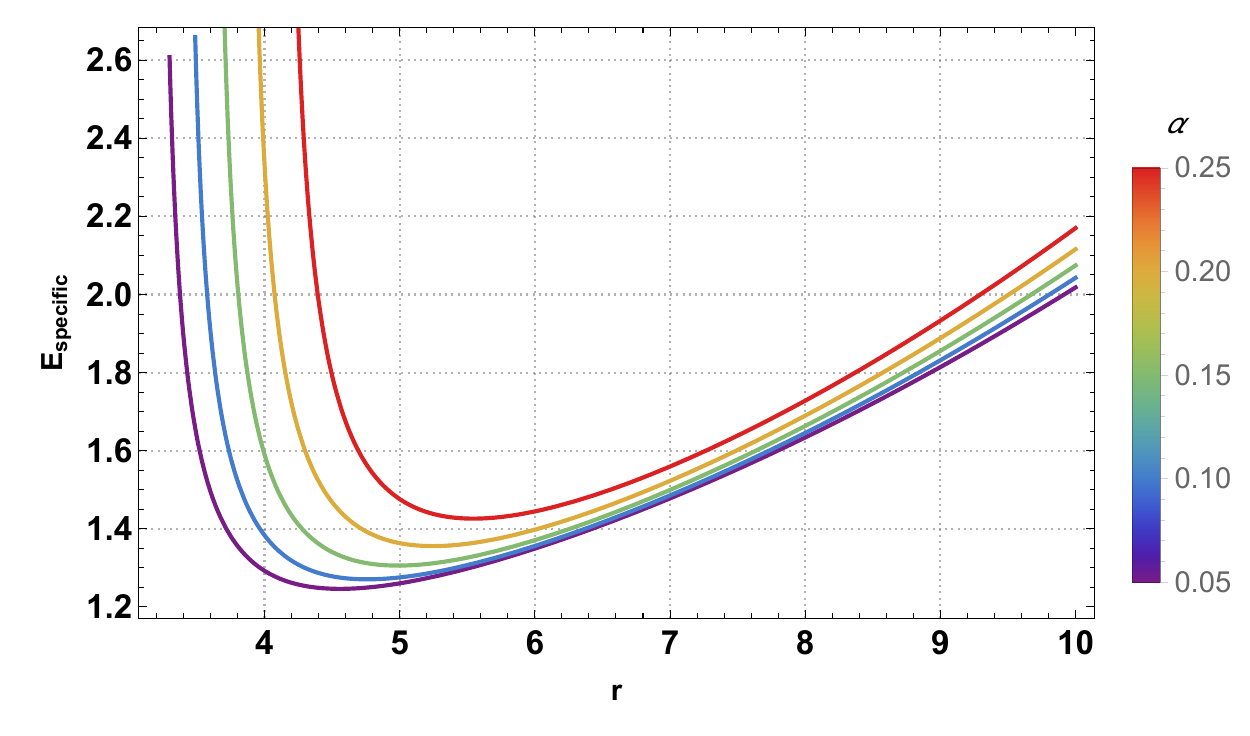}\qquad
    \includegraphics[width=0.45\linewidth]{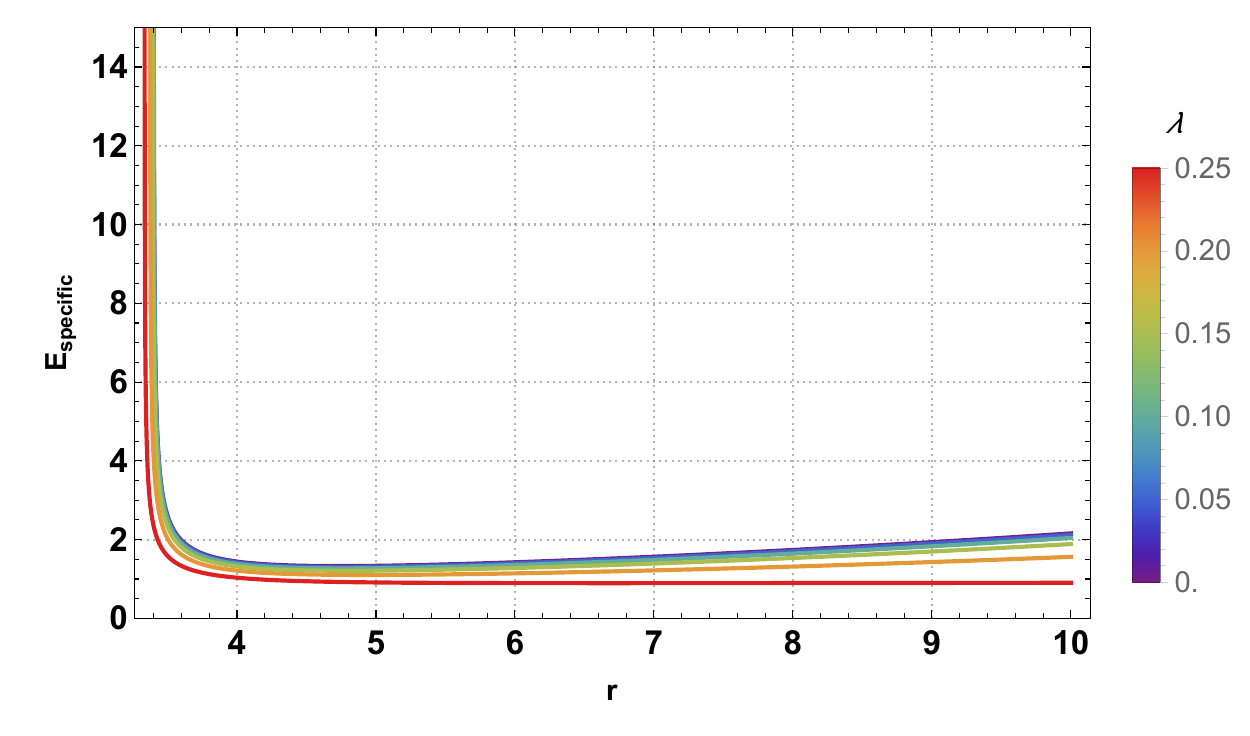}\\
     (a) $\lambda=0.1$ \hspace{8cm} (b) $\alpha=0.1$
    \caption{\footnotesize Behavior of the specific energy $\mathrm{E}(\mbox{specific})$ as a function of $r$ for various values of CoS parameter $\alpha$ and Rastall parameter $\lambda$. Here, $M=1,\,\mathrm{L}=1,\,\mathcal{N}=0.01,\,\Lambda=-0.03$.}
    \label{fig:energy}
\end{figure}

In Figure \ref{fig:angular}, we depict the behavior of the specific angular momentum of test particles showing the effects of CoS parameter $\alpha$ and Rastall parameter $\lambda$, while keeping the QF parameters fixed. In panel (a), we observe that angular momentum increases with increasing $\alpha$. In contrast, this angular momentum decreases with increasing $\lambda$.

Similarly, Figure \ref{fig:energy} illustrate the behavior of the specific energy of test particles by varying CoS parameter $\alpha$ and the Rastall parameter $\lambda$, while keeping the QF parameters fixed. The nature of the specific energy is analogous to the specific angular momentum stated in the previous Figure.

The ISCO radius can be determined using the condition $V''_\text{eff}(r)=0$ which results the following relation of the metric function:
\begin{equation}
    3\,f(r)\, \frac{f'(r)}{r} + f''(r)\, f(r) - 2\, f'(r)^2 = 0.\label{timelike-10}
\end{equation}
Using the metric function $f(r)$ given in Eq.~(\ref{function}), we find the following polynomial equation for ISCO radius $r=r_\text{ISCO}$ as,
\begin{align}
    &3 \left(1 - \alpha - \frac{2\,M}{r} - \frac{\mathcal{N}}{r^m} - \frac{\Lambda}{3} r^2 \right) \left( \frac{2\,M}{r^3} + \frac{m\,\mathcal{N}}{r^{m+2}} - \frac{2\, \Lambda}{3} \right)
+ \left( - \frac{4\,M}{r^3} - \frac{m\, (m+1)\,\mathcal{N}}{r^{m+2}} - \frac{2\,\Lambda}{3} \right) \left( 1 - \alpha - \frac{2\,M}{r} - \frac{\mathcal{N}}{r^m} - \frac{\Lambda}{3}\,r^2 \right)\nonumber\\
&- 2\,\left( \frac{2\,M}{r^2} + \frac{m\,\mathcal{N}}{r^{m+1}} - \frac{2\,\Lambda}{3} r \right)^2 = 0.\label{timelike-11}
\end{align}

\begin{table}[ht!]
\centering
\begin{tabular}{|c|c|c|c|c|}
\hline
$\alpha (\downarrow) \backslash \lambda (\rightarrow )$ & 0.05 & 0.10 & 0.15 & 0.20 \\
\hline
0.05 & 4.53914 & 4.55816 & 4.60239 & 4.73698 \\
\hline
0.12 & 4.82969 & 4.84661 & 4.88659 & 5.01534 \\
\hline
0.15 & 4.97038 & 4.98630 & 5.02397 & 5.14889 \\
\hline
0.20 & 5.23131 & 5.24549 & 5.27861 & 5.39490 \\
\hline
0.25 & 5.53235 & 5.54487 & 5.57243 & 5.67655 \\
\hline
0.30 & 5.88418 & 5.89542 & 5.91639 & 6.00364 \\
\hline
0.35 & 6.30153 & 6.31251 & 6.32594 & 6.39002 \\
\hline
\end{tabular}
\caption{Numerical results of ISCO radius $r=r_\text{ISCO}$ for various values of $\alpha$ and $\lambda$.  Here $M=1,\,\mathcal{N}=0.01,\,\Lambda=-0.03$.}
\label{tab:3}
\end{table}

Equation~(\ref{timelike-11}) is a polynomial in \( r \), for which obtaining an exact analytical solution is highly nontrivial. However, the ISCO radius can be determined numerically by choosing appropriate values for \( M \), \(\mathcal{N}\), \(\lambda\), \( \alpha\) and \(\Lambda\).

In Table \ref{tab:3}, we present numerical values of ISCO radius by varying CoS parameter $\alpha$ and the Rastall parameter $\lambda$, while keeping all other parameters fixed.

\section{Thermodynamic Properties of Black Hole}\label{sec:4}

The geometry of the deformed RN–AdS black hole under consideration is encoded in Eq.~(\ref{function1}). 
To analyze the associated thermodynamic quantities, it is convenient to express the BH mass in terms of the horizon radius $r_{+}$. 
The horizon condition $f(r_{+})=0$ yields
\begin{eqnarray}
1 - \alpha - \frac{2 M}{r_{+}} - \frac{\mathcal{N}}{r_{+}^{m}} - \frac{\Lambda}{3} r_{+}^2=0 ,\label{ss1}
\end{eqnarray}
which can be rearranged to give
\begin{eqnarray}
M=
\frac{r_{+}}{2}\,  \left(
   1   - \alpha  - \frac{\mathcal{N}}{r_{+}^m}
   -\frac{\Lambda}{3}\,r_{+}^{2}
\right).
\label{ss2}
\end{eqnarray}

\begin{figure}[ht!]
    \centering
    \includegraphics[width=0.45\linewidth]{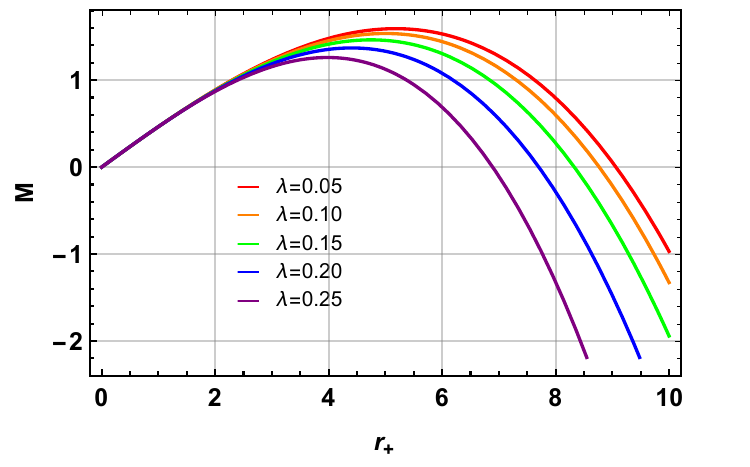}\qquad
    \includegraphics[width=0.45\linewidth]{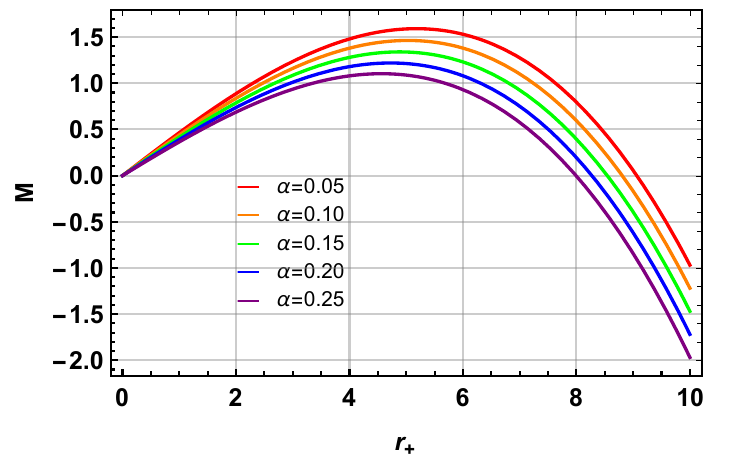}\\
     (a) $\lambda=0.05$ \hspace{8cm} (b) $\alpha=0.05$
    \caption{\footnotesize Behavior of ADS mass $M(r_+)$. Here $\Lambda=-0.03,\,\mathcal{N}=0.01, w=-2/3$.}
    \label{fig1}
\end{figure}

The Hawking temperature follows from the surface gravity relation, 
\begin{eqnarray}
T_{H}=\frac{f'(r_{+})}{4\pi}=
\frac{1}{4\pi\,r_{+}}\Bigg\{
    1 - \alpha+ 
    \dfrac{(m-1)\,\mathcal{N}}{r_{+}^m}-\Lambda r^2_{+}
\Bigg\}.\label{ss3}
\end{eqnarray}

\begin{figure}[ht!]
    \centering
    \includegraphics[width=0.45\linewidth]{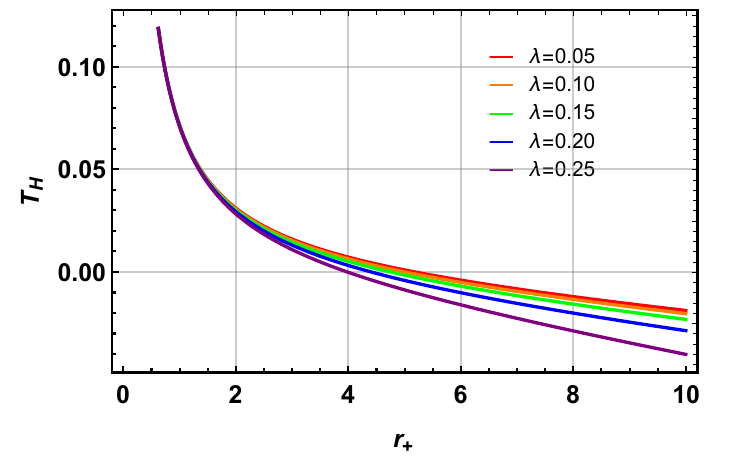}\qquad
    \includegraphics[width=0.45\linewidth]{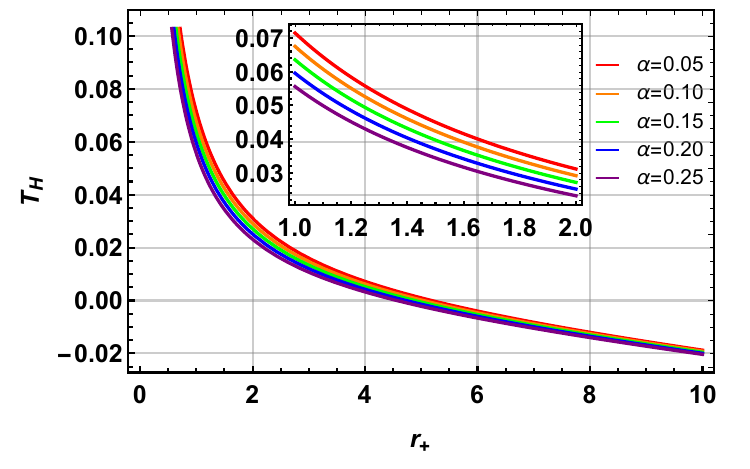}\\
     (a) $\lambda=0.05$ \hspace{8cm} (b) $\alpha=0.05$
    \caption{\footnotesize Behavior of Hawking Temperature. Here $\Lambda=-0.03,\,\mathcal{N}=0.01, w=-2/3$.}
    \label{fig2}
\end{figure}

For static, spherically symmetric configurations, the entropy satisfies the Bekenstein–Hawking area law,
\begin{eqnarray}
S=\frac{A}{4}=\pi r_{+}^{2}.\label{ss4}
\end{eqnarray}
Moreover, in an extended phase space, the thermodynamic pressure is related with the cosmological constant $\Lambda$ as
\begin{equation}
    P=-\frac{\Lambda}{8\pi}.\label{ss5}
\end{equation}

Thereby, the BH mass in terms of entropy and pressure is given by
\begin{equation}
    M=\frac{1}{2}\,\sqrt{\frac{S}{\pi}}\,\left[
   1   - \alpha  -\mathcal{N}\,(\pi/S)^{m/2}
   +\frac{8P}{3}\,S\right].
\label{ss6}
\end{equation}

\subsection{First Law of Thermodynamics}

In BH mass $M$ given in Eq.(\ref{ss6}), considering the cloud of string $\alpha$, the normalization of Quintessence-like field $\mathcal{N}$ as intensive thermodynamic variables, then the BH mass can be expressed as
\begin{equation}
    M=T\,dS+P\,dV+\Theta_{\alpha}\,d\alpha+\Theta_{\mathcal{N}}\,d\mathcal{N},\label{ss7}
\end{equation}
where $\Theta_{\alpha}$ and $\Theta_{\mathcal{N}}$, respectively are the extensive parameters of $\alpha$ and $\mathcal{N}$ and is given by
\begin{align}
    T&=\left(\frac{dM}{dS}\right)_{P,\alpha,\mathcal{N}}=\frac{1}{4\sqrt{\pi S}} \left( 1 - \alpha - \mathcal{N} \left( \frac{\pi}{S} \right)^{m/2} + \frac{8P}{3} S \right)
+
\frac{1}{2} \left( \frac{S}{\pi} \right)^{1/2}
\left[
\frac{8P}{3} + \frac{m \mathcal{N}}{2S} \left( \frac{\pi}{S} \right)^{m/2}
\right],\label{ss8}\\
    \Theta_{\alpha}&=\left(\frac{dM}{d\alpha}\right)_{S,P,\mathcal{N}}=-\frac{1}{2}\,\sqrt{\frac{S}{\pi}},\label{ss9}\\
    \Theta_{\mathcal{N}}&=\left(\frac{dM}{d\mathcal{N}}\right)_{S,P,\alpha}=-\frac{1}{2}\,\left(\frac{S}{\pi}\right)^{(1-m)/2},\label{ss10}\\
\end{align}

The heat capacity at constant pressure can then be written as
\begin{eqnarray}
C_{p}=T_{H}\,\frac{\partial S}{\partial T_{H}}&=& {
    2\pi r_{+} \Bigg\{
        \frac{1 - \alpha}{r_{+}} 
        + \frac{3 \mathcal{N}\, r_{+}^{-3 + \frac{-1 + 3w}{-1 + 3(1+w)\lambda}} \,\Big[w(-1+\lambda)+\lambda\Big]}
               {-1 + 3(1+w)\lambda}
        + \Lambda r_{+}
    \Bigg\}
}\nonumber\\ &&\times{\Bigg\{
     \Lambda-\Big(\frac{-1+\alpha}{r_{+}^{2}}\Big)
    - \frac{
        3 \mathcal{N}\, r_{+}^{-4 + \tfrac{-1 + 3w}{-1 + 3(1+w)\lambda}} \,\Big[w(-1+\lambda)+\lambda\Big]\,\Big[9(1+w)\lambda-2 - 3w\Big]
    }{[1 - 3(1+w)\lambda]^{2}}   
}\Bigg\}^{-1}.
\end{eqnarray}

\begin{figure}[ht!]
    \centering
    \includegraphics[width=0.45\linewidth]{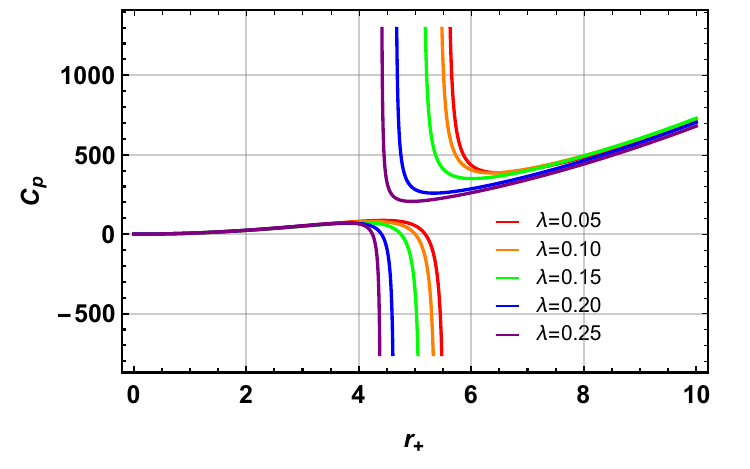}\qquad
    \includegraphics[width=0.45\linewidth]{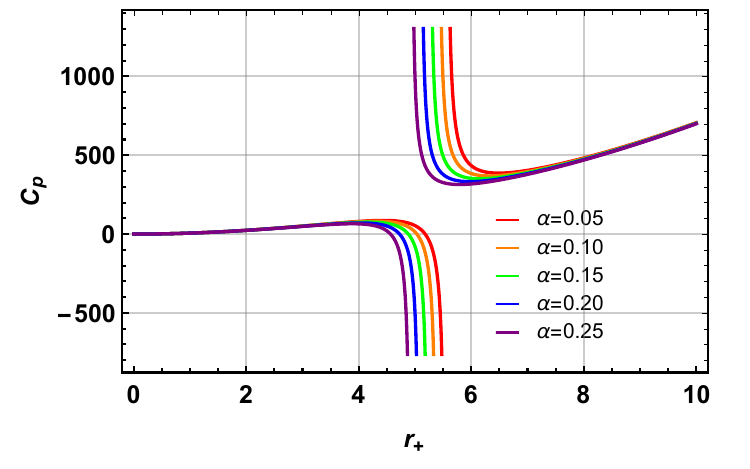}\\
     (a) $\lambda=0.05$ \hspace{8cm} (b) $\alpha=0.05$
    \caption{\footnotesize Behavior of Heat capacity. Here $\Lambda=-0.03,\,\mathcal{N}=0.01, w=-2/3$.}
    \label{fig3}
\end{figure}

Figure~\ref{fig1} illustrates the dependence of the AdS black hole mass $M(r_{+})$ on the horizon radius for different values of the Rastall parameter $\lambda$, Fig.\ref{fig1} (a), and the string cloud parameter $\alpha$, Fig.\ref{fig1} (b). In both cases, the mass exhibits a non-monotonic behavior: it increases with $r_{+}$, attains a maximum at an intermediate scale, and then decreases for large horizon radii due to the dominant negative contribution of the cosmological constant. The plots show that larger values of $\lambda$ or $\alpha$ shift the mass profile downward, reducing both the peak value of the mass and the range of admissible horizon radii for which $M>0$. This behavior highlights the combined role of the Rastall modification and the string cloud background in altering the thermodynamic stability of the black hole.

Figure~\ref{fig2} shows the behavior of the Hawking temperature $T_{H}$ as a function of the horizon radius $r_{+}$ for distinct values of the Rastall parameter $\lambda$, Fig.\ref{fig2} (a), and the string cloud parameter $\alpha$, Fig.\ref{fig2} (b). In both cases, the temperature decreases monotonically with increasing $r_{+}$, approaching negative values for sufficiently large horizons due to the influence of the negative cosmological constant. The inset in panel (b) highlights the near-horizon region, where the temperature is positive and sensitive to variations of $\alpha$. Higher values of $\lambda$ or $\alpha$ suppress the temperature profile, lowering its magnitude and reducing the horizon range for which $T_{H}>0$. This indicates that both the Rastall correction and the string cloud background contribute to destabilizing the thermodynamic behavior of the black hole.

Figure~\ref{fig3} presents the heat capacity $C_{p}$ as a function of the horizon radius $r_{+}$ for different values of the Rastall parameter $\lambda$, Fig.\ref{fig3} (a), and the string cloud parameter $\alpha$, Fig.\ref{fig3} (b). The plots reveal the existence of divergences in $C_{p}$, which signal second-order phase transitions of the black hole. For small horizon radii, the heat capacity is negative, indicating a thermodynamically unstable branch. As $r_{+}$ increases, $C_{p}$ diverges at a critical horizon size, beyond which it becomes positive, corresponding to a stable phase. Increasing either $\lambda$ or $\alpha$ shifts the divergence point toward smaller $r_{+}$ values, reducing the stability range of the black hole. This behavior emphasizes the combined role of Rastall corrections and string cloud effects in controlling the thermodynamic phase structure.

\subsection{Critical Point}

To begin the analysis, it is necessary to specify the thermodynamic equation of state. 
Within the framework of the extended phase space, the cosmological constant $\Lambda$ 
is reinterpreted as a thermodynamic pressure according to  
\begin{equation}
    P = -\frac{\Lambda}{8\pi}. \label{eq:pv1}
\end{equation}
This prescription not only enlarges the thermodynamic description of black holes, 
but also allows the appearance of intricate phase structures and critical phenomena, 
closely resembling those encountered in standard thermodynamic systems, such as 
Van der Waals fluids. Such a reformulation provides the essential groundwork for 
deriving the equation of state and exploring the $P$-$V$ critical behavior. 
In this framework, the pressure can be expressed as  
\begin{equation}
    P = \frac{1}{8\pi r_{+}^{4}}\Bigg[\,4\pi r_{+}^{3}T + r_{+}^{2}(\alpha-1) - 
\dfrac{3 \mathcal{N}\, r_{+}^{\frac{3w-1}{3(1+w)\lambda-1}} \Big(w(\lambda-1)+\lambda\Big)}
{3(1+w)\lambda-1}\Bigg].
\end{equation}

In order to obtain the thermodynamic critical parameters, one needs the expression 
for the thermodynamic volume of the black hole, which takes the form
\begin{equation}
    V = \frac{4\pi r_{+}^{3}}{3}. \label{eq:Vbh}
\end{equation}
Although the direct determination of critical quantities might seem technically involved, 
the procedure can be streamlined by employing the standard method discussed in 
Refs.~\cite{Jafarzade:2017kin,Liu:2016uyd}. In this approach, the critical point is identified by 
imposing the conditions
    ${\partial P}/{\partial V} = 0$, ${\partial^{2} P}/{\partial V^{2}} = 0$,
which ensure the existence of physically acceptable solutions. Solving these 
constraints yields the critical temperature
\begin{equation}
T_c = \frac{1}{2 \pi r_{+}} - \frac{\alpha}{2 \pi r_{+}} 
+ \frac{9 \mathcal{N}\, r_{+}^{\frac{3w-1}{-1+3(1+w)\lambda}-3} 
(1+w)\Big(w(\lambda-1)+\lambda\Big)(4\lambda-1)}
{4 \pi \left[\,1 - 3(1+w)\lambda\,\right]^{2}}.
\end{equation}

\begin{figure}[ht!]
    \centering
    \includegraphics[width=0.45\linewidth]{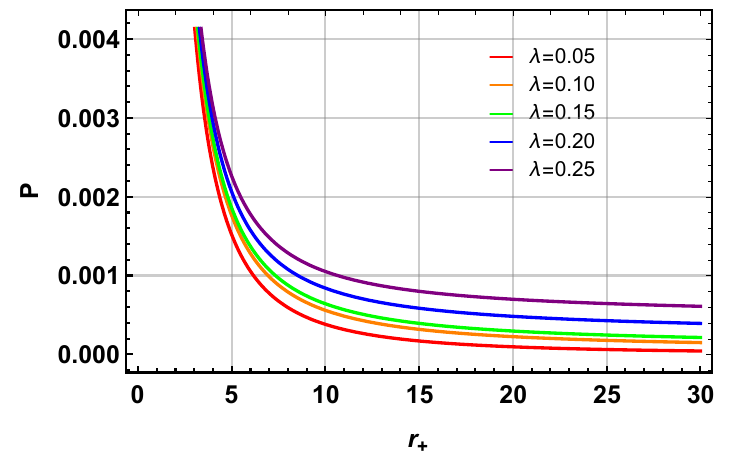}\qquad
    \includegraphics[width=0.45\linewidth]{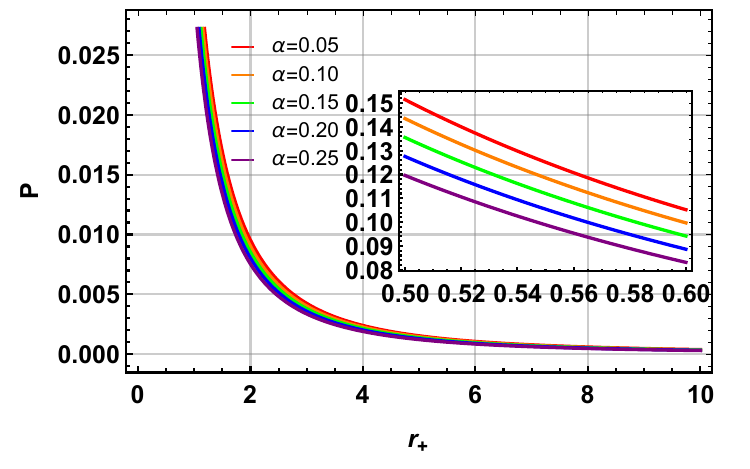}\\
     (a) $\lambda=0.05$ \hspace{8cm} (b) $\alpha=0.05$
    \caption{\footnotesize Behavior of Pressure with $T=T_c$. Here $\mathcal{N}=0.01, w=-2/3$.}
    \label{fig4}
\end{figure}

\begin{figure}[ht!]
    \centering
    \includegraphics[width=0.45\linewidth]{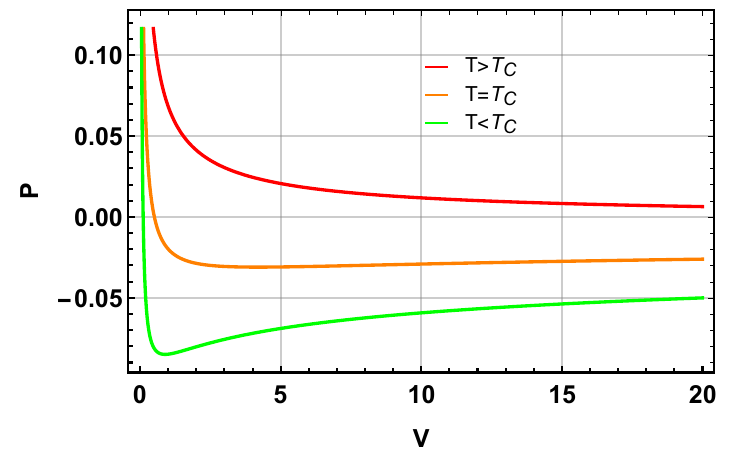}
    \caption{\footnotesize $P$-$V$. Here $\alpha=\lambda=0.05,\,\mathcal{N}=0.01, w=-2/3$.}
    \label{fig5}
\end{figure}

Figure~\ref{fig4} illustrates the behavior of the pressure $P$ as a function of the horizon 
radius $r_{+}$ for different values of the Rastall parameter $\lambda$ (Fig.\ref{fig4}(a)) and the 
string cloud parameter $\alpha$ (Fig.\ref{fig4} (b)). As shown in Fig.\ref{fig4}(a), increasing $\lambda$ leads to a systematic upward shift in the pressure curves, particularly at larger values of $r_{+}$, indicating that the Rastall correction enhances the effective pressure of the system. 
Fig.\ref{fig4} (b) displays the variation of $P(r_{+})$ with different values of $\alpha$ at fixed 
$\lambda=0.05$. In this case, higher values of $\alpha$ reduce the pressure for a given 
horizon radius, as highlighted in the inset, which zooms into the small-$r_{+}$ region. 
These results demonstrate the nontrivial role played by $\lambda$ and $\alpha$ in shaping 
the $P$-$V$ thermodynamics and in determining the location of possible critical points.

Figure~\ref{fig5} shows the $P$-$V$ isotherms for different temperature regimes relative 
to the critical temperature $T_{C}$, with fixed parameters 
$\alpha = \lambda = 0.05$, $\mathcal{N}=0.01$, and $w=-2/3$. 
The red curve corresponds to the case $T>T_{C}$, where the pressure decreases monotonically 
with increasing thermodynamic volume. The orange curve illustrates the critical isotherm 
$T=T_{C}$, marking the onset of the phase transition. Finally, the green curve depicts 
the case $T<T_{C}$, where the characteristic Van der Waals-like behavior is evident: 
a local minimum and maximum appear, signaling the coexistence of small and large black hole 
phases. This behavior confirms the analogy between the thermodynamics of the system 
and that of standard fluid models, such as the Van der Waals fluid.

\subsection{Phase Transition}

Finally, the Gibbs free energy is defined by
\begin{eqnarray}
G=M-T_{H}S=\frac{\mathcal{N}\, r^{-\frac{3\left[w(\lambda -1)+\lambda\right]}{-1+3(1+w)\lambda}}
\left(1+3w-6(1+w)\lambda\right)}{6(1+w)\lambda-2}
- \frac{r^{3}\Lambda}{3}.
\end{eqnarray}

\begin{figure}[ht!]
    \centering
    \includegraphics[width=0.45\linewidth]{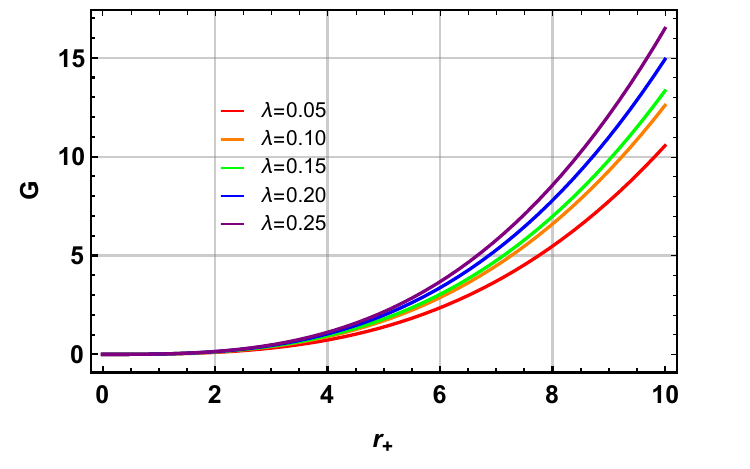}
    \caption{\footnotesize Behavior of Gibbs free energy. Here $\Lambda=-0.03,\,\mathcal{N}=0.01, w=-2/3$.}
    \label{fig6}
\end{figure}

Figure~\ref{fig6} depicts the behavior of the Gibbs free energy $G$ as a function of the horizon radius $r_{+}$ for the considered AdS black hole surrounded by a cloud of strings and a quintessence field in the Rastall framework. The plot clearly exhibits a characteristic swallowtail structure, which is a hallmark of a first-order phase transition between small and large black hole phases. As the horizon radius increases, the system undergoes a discontinuous change in Gibbs free energy, indicating the coexistence of two thermodynamically distinct phases at the critical temperature. This behavior confirms the Van der Waals-like nature of the black hole thermodynamics and highlights the significant influence of both the Rastall parameter $\lambda$ and the string cloud parameter $\alpha$ on the phase structure, modifying the location and extent of the phase transition region.

\section{Conclusions}\label{sec:5}

In this study, we tested the properties of static AdS BHs within the Rastall gravity framework, taking into account the effects of surrounding string clouds and quintessence fields, thereby extending the standard general relativistic understanding of black hole dynamics and thermodynamics. The analysis began with the construction of the modified spacetime geometry, where the presence of the Rastall parameter \( \lambda \), the string cloud parameter \( \alpha \), and the quintessence normalization constant \( N \) introduced nontrivial deviations from Einstein’s theory, directly affecting the horizon structure and the effective potential governing particle motion. 

We carried out a detailed study of null geodesics, highlighting the impact of these modifications on photon trajectories, the photon sphere radius, the black hole shadow size, the associated effective radial force, and the topological features of photon rings. It was shown that both \( \alpha \) and \( \lambda \) played crucial roles in shifting the photon sphere and, consequently, altering the observable shadow profile-an effect of significant relevance for astrophysical observations associated with the Event Horizon Telescope and gravitational lensing phenomena. Extending the investigation to timelike geodesics, we analyzed the motion of massive particles and computed the conditions for innermost stable circular orbits (ISCOs). The results demonstrated that the specific energy and angular momentum of test particles, as well as the ISCO radius, depended strongly on the interplay between Rastall corrections, string clouds, and quintessence, providing valuable insights into the potential modifications of accretion disk dynamics in alternative theories of gravity.

On the thermodynamic side, we derived expressions for the black hole mass, Hawking temperature, entropy, and heat capacity, and established their dependence on the chosen parameters. In particular, we showed how thermodynamic stability and the occurrence of phase transitions were affected by non-Einsteinian contributions. 

Overall, our results illustrated that Rastall gravity, when coupled with external matter distributions such as string clouds and quintessence, introduced significant modifications to both the geodesic structure and thermal behavior of BH. This provided a consistent platform for bridging gravitational physics, quantum field theory, and statistical mechanics. Furthermore, the study confirmed that the Rastall framework, enriched with physically motivated sources, offers fertile ground for exploring the deep connections between geometry, dynamics, and thermodynamics in asymptotically AdS spacetimes.

Although the present work has clarified several key aspects of how Rastall gravity, string clouds, and quintessence influence black hole physics, there remain important directions for future investigation and potential observational testing. A natural extension of this study would involve generalizing the analysis to rotating BH within the Rastall framework. In such a setting, the combined effects of angular momentum, string cloud distributions, and quintessence fields could lead to substantial modifications in the geodesic structure, the ergoregion, and phenomena such as superradiant scattering. 

Another promising direction would be the study of quasinormal modes and the stability of linear perturbations in these extended backgrounds. These modes are directly linked to gravitational wave signals and could offer novel observational signatures to constrain the Rastall parameter. From a thermodynamic perspective, future work could also explore the thermodynamic topology in an extended phase space formalism, in which the cosmological constant is interpreted as thermodynamic pressure. This enables us the thermodynamic behavior of BH under Rastall gravity corrections.

{\small
\section*{Acknowledgments}

F.A. acknowledges the Inter University Centre for Astronomy and Astrophysics (IUCAA), Pune, India for granting visiting associateship.

\section*{Data Availability Statement}

This manuscript has no associated data.

\section*{Conflict of Interests}

Author declare (s) no conflict of interest.
}

\end{document}